\RequirePackage{ifpdf}
\documentclass[letterpaper]{JHEP3}
\usepackage{amsmath}
\usepackage{epsfig}
\usepackage{subfigure}

\DeclareSymbolFont{matha}{OML}{txmi}{m}{it}
\DeclareMathSymbol{\varv}{\mathord}{matha}{118}

\newcommand{\roughly}[1]{\mathrel{\raise.3ex\hbox{$#1$\kern-0.85em
\lower1ex\hbox{$\sim$}}}}

\newcommand{\lsim}{\roughly<}
\newcommand{\gsim}{\roughly>}

\def\endignore{}
\def\ignore #1\endignore{} 

\def\la{{\bigl \langle}}
\def\ra{{\bigr \rangle}}
\def\cA{{\cal A}}

\def\cQ{{\cal Q}}

\def\cX{{\cal X}}

\def\cL{{\cal L}}

\def\cO{{\cal O}}

\def\cG{{\cal G}}
\def\cY{{\cal Y}}
\def\cR{{\cal R}}
\def\cT{{\cal T}}

\def\cZ{{\cal Z}}

\def\Bext{{B_{\rm ext}}}

\def\BLF{{\scriptscriptstyle BLF}}

\newbox\charbox
\newbox\slabox
\def\slsh#1{{      
        \setbox\charbox=\hbox{$#1$}
        \setbox\slabox=\hbox{$/$}
        \dimen\charbox=\ht\slabox
        \advance\dimen\charbox by -\dp\slabox
        \advance\dimen\charbox by -\ht\charbox
        \advance\dimen\charbox by \dp\charbox
        \divide\dimen\charbox by 2
        \raise-\dimen\charbox\hbox to \wd\charbox{\hss/\hss}
        \llap{$#1$}
}}

\def\exd{{\hbox{d}}}
\def\d{\exd}

\def\bea{\begin{eqnarray}}
\def\eea{\end{eqnarray}}
\def\be{\begin{equation}}
\def\ee{\end{equation}}

\def\ssA{{\scriptscriptstyle A}}
\def\ssB{{\scriptscriptstyle B}}
\def\ssC{{\scriptscriptstyle C}}
\def\ssD{{\scriptscriptstyle D}}

\def\ssF{{\scriptscriptstyle F}}

\def\ssM{{\scriptscriptstyle M}}
\def\ssN{{\scriptscriptstyle N}}
\def\ssP{{\scriptscriptstyle P}}
\def\ssQ{{\scriptscriptstyle Q}}
\def\ssR{{\scriptscriptstyle R}}

\def\ssT{{\scriptscriptstyle T}}

\def\ssV{{\scriptscriptstyle V}}

\def\ssZ{{\scriptscriptstyle Z}}

\def\EH{{\scriptscriptstyle EH}}

\def\nn{\nonumber}

\def\d{\mathrm{d}}

\def\({\left(}
\def\){\right)}

\def\pref#1{(\ref{#1})}
\def\eff{{\rm eff}}

\def\gR{g_\ssR}

\def\d{\mathrm{d}}

\numberwithin{equation}{section}

\title{EFT for Vortices with Dilaton-dependent Localized Flux}

\author{C.P.~Burgess,$^{1,2,3}$ Ross~Diener${}^{1,2}$ and M. Williams$^{4}$ \\
$^1$ Physics \& Astronomy, McMaster University,
Hamilton, ON, Canada, L8S 4M1\\
$^2$ Perimeter Institute for Theoretical Physics, Waterloo, ON, Canada N2L 2Y5\\
${}^3$ Division PH\,-TH, CERN, CH-1211, Gen\`eve 23, Suisse\\
$^4$ Instituut voor Theoretische Fysica, KU Leuven,
B-3001 Leuven, Belgium
}

\preprint{Preprint: CERN-PH-TH-2015-054}

\date{\today}

\abstract { We study how codimension-two objects like vortices back-react gravitationally with their environment in theories (such as 4D or higher-dimensional supergravity) where the bulk is described by a dilaton-Maxwell-Einstein system. We do so both in the full theory, for which the vortex is an explicit classical `fat brane' solution, and in the effective theory of `point branes' appropriate when the vortices are much smaller than the scales of interest for their back-reaction (such as the transverse Kaluza-Klein scale). We extend the standard Nambu-Goto description to include the physics of flux-localization wherein the ambient flux of the external Maxwell field becomes partially localized to the vortex, generalizing the results of a companion paper \cite{Companion} to include dilaton-dependence for the tension and localized flux. In the effective theory, such flux-localization is described by the next-to-leading effective interaction, and the boundary conditions to which it gives rise are known to play an important role in how (and whether) the vortex causes supersymmetry to break in the bulk. We track how both tension and localized flux determine the curvature of the space-filling dimensions. Our calculations provide the tools required for computing how scale-breaking vortex interactions can stabilize the extra-dimensional size by lifting the dilaton's flat direction. For small vortices we derive a simple relation between the near-vortex boundary conditions of bulk fields as a function of the tension and localized flux in the vortex action that provides the most efficient means for calculating how physical vortices mutually interact without requiring a complete construction of their internal structure. In passing we show why a common procedure for doing so using a $\delta$-function can lead to incorrect results. Our procedures generalize straightforwardly to general co-dimension objects. }

\begin{document}

\section{Introduction}

A `vanilla' Nielsen-Olesen vortex \cite{NOSolns,CStrings} carries $U(1)$ flux but this flux is normally expelled from the region outside of the vortex. In this paper we study the gravitational response of vortices (or `fat' branes) that carry localized amounts of an external magnetic flux that is {\em not} expelled from its surrounding environment (so-called `Dark Vortices' or `Dark Strings' \cite{DStrings}). Our description of these systems closely parallels our companion paper \cite{Companion}, extending it to the case where effective couplings are functions of the bulk dilaton that tends to appear in supersymmetric theories.

Vortices which partially localize bulk flux can arise within supersymmetric theories in various dimensions, and whether or not their presence breaks supersymmetry depends on the relative size of their tension and the amount localized bulk flux they carry \cite{AccidentalSUSY}. Because of this their tension and localized flux compete with one another in the amount of curvature produced by their back-reaction on their surrounding geometry. Our main goal is to explore this competition in detail and to identify precisely how it depends on the various parameters that describe the vortex physics.

We have several purposes in mind when doing so. First and foremost we wish to understand how brane back-reaction influences the transverse geometry through which the vortices move, and in particular how they stabilize the value of the dilaton and so set the size of the transverse dimensions and the curvature of the geometry induced on their world-sheet. Much is known about the systems in the limit when the sources are pointlike \cite{BLFFluxQ}, and in particular it is known that the world-sheet geometries are exactly flat (at the classical level) if the dilaton should have a vanishing derivative at all source positions \cite{ScaleLzero} (a result which we also re-derive here).\footnote{Indeed this underlies the study of these system as potential approaches \cite{SLED, LesHouches} to the cosmological constant problem \cite{LesHouches,CCprob}.} What this leaves open is whether there exists any kind of source for which a vanishing near-source dilaton derivative is possible and, even if so, whether the resulting curvature is then nonzero but dominated by finite-size vortex effects that could be suppressed for small vortices but not vanishing.

We find three main results.
\begin{itemize}
\item {\em Vortex-dilaton coupling:} In general the radial derivative of any bulk field very near a point source is completely controlled by the derivative of the source action with respect to the field of interest  \cite{BLFFluxQ, 6DSUSYUVCaps, UVCaps} (much like the quantity $\lim_{r\to 0} r^2 \partial V/\partial r$ is dictated by a point-source's charge, $Q \propto \delta S/\delta V$ in electrostatics). So naively a vanishing dilaton derivative at a the position of a source brane is arranged by not coupling the dilaton to the brane at all. While we confirm the truth of this assertion, we also find it is harder to completely avoid such a brane coupling to the dilaton than one might think. More specifically, we find that although it {\em is} possible to arrange a dilaton-free tension, it is much more difficult to arrange dilaton-free localized flux. It is more difficult because in a supersymmetric bulk the value of the dilaton sets the local size of the gauge coupling for the flux.
\item {\em Modulus stabilization:} By computing how vortex-bulk energetics depend on the value of the dilaton we verify earlier claims \cite{BLFFluxQ,6DSUSYUVCaps,6DHiggsStab} that (with two transverse dimensions) brane couplings generically stabilize the size of the transverse dimensions in supersymmetric models, in a manner similar to Goldberger-Wise stabilization \cite{GoldWis} in 5D. They do so because they break the classical scale invariance of the bulk supergravity that prevents the bulk from stabilizing on its own (through {\em eg}\, flux stabilization). The tools we provide allow an explicit calculation of the energetics as a function of the dilaton field (and so in principle allow a calculation of the stabilizing dilaton potential).
\item {\em Low-energy on-brane curvature:} We find that the same dynamics usually also curves the dimensions along the vortex world-sheets, and generically does so by an amount commensurate with their tension, $R \sim G_\ssN \check T$, where $\check T$ is the vortex tension (defined more precisely below) and $G_\ssN$ is Newton's constant for observers living on the brane. For specific parameter regimes the on-vortex curvature can be less than this however, being suppressed by the deviation of the vortex from scale invariance (when this is small) and the ratio of the vortex size to the
\item{\em Matching and effective descriptions:} We describe our analysis throughout in two complimentary ways. On one hand we do so using the full (UV) description within which the vortices are explicit classical solutions. We then do so again using the lower-energy (IR) extra-dimensional effective theory within which the vortices are regarded as point sources because their sizes are not resolved. By comparing these calculations we provide explicit matching formulae that relate dilaton-dependent effective parameters (like tension and localized flux) to underlying properties of the UV completion.
\item{\em Efficient description of point-source back-reaction:} We provide explicit formula that relate the near-source boundary conditions of bulk fields in terms of the source tension and localized flux. Because these boundary conditions determine the integration constants of the external bulk solutions they efficiently solve the back-reaction problem in a way that does not depend on providing a detailed construction of the internal microstructure of the sources. As such they provide the most efficient way to describe how such sources gravitate, and the framework within which to renormalize the divergences associated with the singularity of bulk fields at the source positions \cite{6DHiggsStab,ClassRenorm}. We show in passing why a commonly used $\delta$-function way of trying to infer brane-bulk interactions can give incorrect results.
\end{itemize}

Our conclusions also include several more technical observations about the gravitational physics of small localized brane sources. In particular, we identify how stress-energy conservation strongly constrains the components of the source source stress energy tensor, allowing in particular the extra-dimensional off-brane stress-energy components to be dictated in terms of the effective action governing the on-brane degrees of freedom. We do so by using the vortex system to explicitly construct the effective theory of point sources describing the extra-dimensional response to the vortices on scales too long to resolve the vortex size. In particular we show how the radial Einstein `constraint' determines the two nontrivial off-brane components of stress energy (for rotationally invariant -- monopole -- sources) on scales much longer than the size of the vortices themselves.

\subsubsection*{A road map}

We organize our discussion as follows. The following section, \S\ref{sec:bulksystem}, describes the bulk system in the absence of any localized sources. The purposes of doing so is to show how properties of the bulk physics (such as extra-dimensional size and on-brane curvature) are constrained by the field equations, which controls the extent to which they depend on the properties of any source branes.

Then come sources. \S\ref{sec:sourcesystem} first describes the source physics in terms of a UV theory within which the branes can be found as explicit classical vortex solutions. This is followed, in \S\ref{sec:6DIR} by a discussion of the higher-dimensional effective theory that applies on length scales too large to resolve the vortices, but small enough to describe the extra-dimensional geometry. In this regime the vortices are described by effective point sources, and we find their properties as functions of the choices made in the full UV theory. \S\ref{sec:gravresponse} summarizes how the main physical quantities scale as functions of the couplings assumed between the vortex and the dilaton.
Our conclusions are summarized in a final discussion section, \S\ref{section:discussion}.

\section{The Bulk}
\label{sec:bulksystem}

We start by outlining the action and field equations of the bulk, which we take to be an Einstein-Maxwell-scalar system. Our goal is to understand how bulk properties (such as curvatures) are related to the asymptotic behaviour of the fields near any localized sources. We return in later sections describing the local sources in terms of vortices, with the goal of understanding what features of the source control the near-source asymptotics. We imagine the bulk to span $D = d+2$ spacetime dimensions with the $d$-dimensional sources localized in two transverse dimensions. The most interesting cases of practical interest are the cosmic string [with $(D,d) = (4,2)$] and the brane-world picture [with $(D,d) = (6,4)$]. Since this section on the bulk duplicates standard results \cite{ScaleLzero, SLED, GGP,  DistributedSUSY}, aficionados in a hurry should skip it and go directly to the next section.

\subsection{Action and field equations}
\label{subsec:actionFE}

The bulk action of interest is given by
\bea \label{SB}
 S_\ssB &=& - \int \exd^{d+2}x \; \sqrt{-g} \left[ \frac{1}{2\kappa^2} \; g^{\ssM\ssN} \Bigl( \cR_{\ssM \ssN} + \partial_\ssM \phi \, \partial_\ssN \phi \Bigr) + V_\ssB(\phi)  + \frac14  e^{-\phi}  A_{\ssM \ssN} A^{\ssM \ssN}  \right] \nn\\
 &=:& - \int \exd^{d+2}x \; \sqrt{-g} \; \Bigl( L_\EH + L_\phi + L_\ssA \Bigr)
 =: - \int \exd^{d+2}x \; \sqrt{-g} \;  L_\ssB
\eea
where\footnote{We use Weinberg's curvature conventions \cite{Wbg}, which differ from those of MTW \cite{MTW} only by an overall sign in the definition of the Riemann tensor.} $A_{\ssM \ssN} = \partial_\ssM A_\ssN - \partial_\ssN A_\ssM$ is a $D$-dimensional gauge field strength and $\cR_{\ssM\ssN}$ denotes the $D$-dimensional Ricci tensor. The second line defines the Einstein-Hilbert, scalar and gauge contributions --- {\em i.e.} $L_\EH$, $L_\phi$ and $L_\ssA$ --- in terms of the items in the line above, with $L_\ssB$ denoting their sum. When needed explicitly we take the scalar potential to be
\be
 V_\ssB(\phi) = V_0\,  e^\phi \,,
\ee
and in the special case $V_0 = 2g_\ssR^2/\kappa^4$ this corresponds to a subset of the action for Nishino-Sezgin supergravity \cite{NS} when $d = 4$ (so $D = 6$). In this case $g_\ssR$ is the gauge coupling constant for a specific gauged $U(1)_\ssR$ symmetry that does not commute with 6D supersymmetry. In what follows we denote by $g_\ssA$ the gauge coupling for the gauge field $A_\ssM$, although $g_\ssA = g_\ssR$ in the most interesting\footnote{Enhanced interest comes from the unbroken 4D supersymmetry that these configurations can enjoy.} situation where this gauge field is the $U(1)_\ssR$ gauge field.

\subsubsection*{Scaling properties}

For later purposes notice that $L_\ssB$ scales homogeneously, $L_\ssB \to s^{-1} L_\ssB$ under the rigid rescalings $g_{\ssM \ssN} \to s \; g_{\ssM \ssN}$ and $e^\phi \to s^{-1} e^\phi$, which as a consequence is a symmetry of the classical equations of motion. The corresponding transformation for the action is $S_\ssB \to s^{d/2} S_\ssB$, as is most easily seen by transforming to a scale-invariant metric $\hat g_{\ssM\ssN} = e^{\phi} g_{\ssM\ssN}$, in which case all terms of the bulk lagrangian are proportional to $e^{-d\phi/2}$ with $\phi$ otherwise only appearing through its derivative, $\partial_\ssM \phi$. Besides ensuring classical scale invariance this also shows that it is the quantity $e^{d\phi/2}$ that plays the role of $\hbar$ in counting loops within the bulk part of the theory.

The bulk system enjoys a second useful scaling property. If we rescale the gauge field so $\cA_\ssM := g_\ssA A_\ssM$ then arbitrary constant shifts $\phi \to \phi + \phi_\star$ leave the action unchanged provided we also rescale the couplings by $g_\ssA \to g_{\ssA\star} := g_\ssA e^{\phi_\star/2}$ and $V_0 \to V_\star := V_0 \, e^{\phi_\star}$ (or, if $V_0 = 2 g_\ssR^2 / \kappa^4$, equivalently $g_\ssR \to g_{\ssR\star} := g_\ssR \, e^{\phi_\star/2}$). This is convenient inasmuch as $\phi = 0$ can always be chosen to be the present-day vacuum provided the values of constants like $g_\ssA$ and $V_0$ are chosen appropriately.

\subsubsection*{Bulk field equations}

The field equations for the Maxwell field arising from the bulk action are
\be \label{BAeq}
 \frac{1}{\sqrt{-g}} \, \partial_\ssM \Bigl( \sqrt{-g} \; e^{-\phi} A^{\ssM \ssN} \Bigr) = 0 \,,
\ee
which is supplemented (as usual) by the Bianchi identity $\exd A_{(2)} = 0$, for the 2-form $A_{\ssM\ssN}$. The bulk dilaton equation is similarly
\be \label{Bdilatoneq}
 \Box \phi = \frac{1}{\sqrt{-g}} \, \partial_\ssM \Bigl( \sqrt{-g} \; g^{\ssM\ssN} \partial_\ssN \phi \Bigr) = \kappa^2 \Bigl( V_\ssB - L_\ssA \Bigr) \,,
\ee
while the Einstein equations can be written in their trace-reversed form
\be \label{TrRevEin}
 \cR_{\ssM \ssN} = - \kappa^2 X_{\ssM \ssN} \,,
\ee
where $X_{\ssM \ssN} := T_{\ssM \ssN} - (1/d)\, g_{\ssM \ssN} \, {T^\ssP}_\ssP$ and the bulk stress-energy tensor is
\be
 T^{(\ssB)}_{\ssM\ssN} = \frac{1}{\kappa^2} \, \partial_\ssM \phi \, \partial_\ssN \phi + e^{-\phi} A_{\ssM\ssP} {A_\ssN}^\ssP - g_{\ssM \ssN} \left[ \frac{(\partial \phi)^2}{2\kappa^2} + V_{\ssB} + L_{\ssA} \right] \,.
\ee
Notice that it is a special feature of the split into $D = d+2$ dimensions that any maximally symmetric contribution to the $d$-dimensional part of the stress-energy, $T_{\mu\nu} \propto g_{\mu\nu}$, drops out of $X_{\mu\nu}$ and so naively does not contribute directly to the 4D Ricci curvature, $\cR_{\mu\nu}$.

\subsection{Symmetry ans\"atze}
\label{subsec:ansatz}

Our interest throughout is in solutions that are maximally symmetric in the $d$ `spacetime' (or `on-brane') dimensions (spanned by $x^\mu$) and axially symmetric in the 2 `transverse' dimensions (spanned by $y^m$). We therefore specialize to fields that depend only on the proper distance, $\rho$, from the points of axial symmetry, and assume the only nonzero components of the gauge field strength, $A_{mn}$, lie in the transverse two directions. We choose the metric to be of the general warped-product form
\be \label{productmetric}
 \exd s^2 = g_{\ssM \ssN} \, \exd x^\ssM \exd x^\ssN = g_{mn} \, \exd y^m \exd y^n + g_{\mu\nu} \, \exd x^\mu \exd x^\nu \,,
\ee
with
\be \label{warpedprod}
 g_{mn} = g_{mn}(y) \qquad \hbox{and} \qquad
 g_{\mu\nu}(x,y)  =  W^2(y) \, \check g_{\mu\nu}(x) \,,
\ee
where $\check g_{\mu\nu}(x)$ is the maximally symmetric metric on $d$-dimensional de Sitter, Minkowski or anti-de Sitter space. The corresponding Ricci tensor is $\cR_{\ssM \ssN} \, \exd x^\ssM \exd x^\ssN = \cR_{\mu\nu} \, \exd x^\mu \exd x^\nu + \cR_{mn} \, \exd y^m \exd y^n$, and is related to the Ricci curvatures, $\check R_{\mu\nu}$ and $R_{mn}$, of the metrics $\check g_{\mu\nu}$ and $g_{mn}$ by
\be
 \cR_{\mu\nu} = \check R_{\mu\nu} + g^{mn} \Bigl[ (d-1) \partial_m W \partial_n W + W \nabla_m \nabla_n W \Bigr] \, \check g_{\mu\nu} \,,
\ee
and
\be \label{cR2vsR2}
 \cR_{mn} = R_{mn} + \frac{d}{W} \; \nabla_m  \nabla_n W \,,
\ee
where $\nabla$ is the 2D covariant derivative built from $g_{mn}$. For the axially symmetric 2D metrics of interest we make the coordinate choice
\be \label{xdmetric}
  g_{mn} \, \exd y^m \exd y^n = A^2(r) \, \exd r^2 + B^2(r) \, \exd \theta^2  =  \exd \rho^2 + B^2(\rho) \, \exd \theta^2 \,,
\ee
where proper distance, $\rho$, is defined by $\exd \rho = A(r) \exd r$. Some useful properties of these geometries are given in \cite{Companion}.

With these choices the field equation simplify to coupled nonlinear ordinary differential equations. Denoting differentiation with respect to proper distance, $\rho$, by primes, the gauge field equations become
\be \label{BAeom}
  \left( \frac{ e^{-\phi} W^d A_{\rho\theta}}{B} \right)' = 0 \,,
\ee
while the field equation for the dilaton becomes
\be \label{Bdilatoneom2}
 \frac{1}{BW^d} \, \Bigl( BW^d \, \phi' \Bigr)' = \kappa^2 \Bigl(V_\ssB - L_\ssA \Bigr) =: \kappa^2 \, \cX_\ssB \,,
\ee
where the last equality defines $\cX_\ssB := V_\ssB - L_{\ssA}$. With the assumed symmetries the nontrivial components of the matter stress-energy are
\be \label{Tmunusymform}
 T_{\mu\nu}  = - g_{\mu\nu} \; \varrho_\ssB \,, \qquad
 {T^\rho}_\rho = \cZ - \cX  \qquad \hbox{and} \qquad
 {T^\theta}_\theta = -( \cZ + \cX )  \,,
\ee
where the bulk contribution to $\cX$ is $\cX_\ssB$ as defined above and
\be
 \varrho_{\ssB} := \frac{(\phi')^2}{2\kappa^2} + V_\ssB + L_{\ssA} \qquad \hbox{and} \qquad
 \cZ_\ssB := \frac{(\phi')^2}{2\kappa^2} \,.
\ee

The three independent components of the trace-reversed bulk Einstein equations then consist of those in the directions of the $d$-dimensional on-brane geometry,
\be \label{B4DTrRevEin}
 \cR_{\mu\nu} = - \kappa^2 X_{\mu\nu}
 = - \frac{2}{d} \; \kappa^2 \cX_\ssB \; g_{\mu\nu}  \,,
\ee
of which maximal symmetry implies the only nontrivial combination is the trace
\be \label{BavR4-v1}
 \cR_{(d)} := g^{\mu\nu} \cR_{\mu\nu} = \frac{\check R}{W^2} + \frac{d}{BW^d} \Bigl( BW' W^{d-1} \Bigr)' = -2\kappa^2 \cX_\ssB \,,
\ee
and we use the explicit expression for $\cR_{(d)}$ in terms of $\check R$ and $W$.

The components dictating the 2-dimensional transverse geometry similarly are $\cR_{mn} = - \kappa^2 X_{mn}$, which for the bulk has the following two nontrivial components:
\be \label{BavR2}
 \cR_{(2)} := g^{mn} \cR_{mn} = R + d \left( \frac{W''}{W} + \frac{B'W'}{BW} \right) = - \kappa^2 {X^m}_m = - 2 \kappa^2 \left[  \varrho_{\ssB} - \left( 1 - \frac{2}{d} \right) \cX_\ssB \right] \,,
\ee
and the other can be the difference between its two diagonal elements
\be
 {\cG^\rho}_\rho - {\cG^\theta}_\theta = {\cR^\rho}_\rho - {\cR^\theta}_\theta = - \kappa^2 \left( {T^\rho}_\rho - {T^\theta}_\theta \right) \,,
\ee
which for the assumed geometry becomes
\be \label{BnewEinstein}
 \frac{B}{W} \left( \frac{W'}{B} \right)' = - \frac{2}{d} \; \kappa^2 \cZ_\ssB = - \frac{(\phi')^2}{d} \le 0 \,.
\ee
This shows that $W'/B$ is a monotonically decreasing function of $\rho$.

A useful linear combination of the above Einstein equations corresponds to the $(\theta\theta)$ component of the trace-reversed equation, which for the bulk reads
\be \label{BXthetathetaeEinstein}
 \frac{ (B' W^d)' }{BW^d} = - \kappa^2  \left[ \varrho_{\ssB} - \cZ_\ssB - \left( 1 - \frac{2}{d} \right) \cX_\ssB \right] = - 2\kappa^2 \left( L_{\ssA} + \frac{\cX_\ssB}{d} \right)  \,.
\ee

\subsection{Bulk solutions}
\label{ssec:bulksolns}

We next describe some of the properties of the solutions to these field equations. In order to accommodate our later inclusion of the equations governing any localized sources, we examine solutions of the bulk equations only within a domain, the `bulk': $\Bext$, which consists of the full 2D geometry transverse to the sources from which small volumes (`Gaussian pillboxes', $X_\varv$) are excised. These pillboxes completely enclose any sources that might be present. In practice we define $\Bext$ such that the radial proper distance coordinate satisfies $\rho_- \le \rho \le \rho_+$, with the sources lying just outside of this range. When not specifying which source is of interest, we generically use $\rho_\varv = \{ \rho_+, \rho_- \}$ to denote the boundary between the source and the bulk.

\subsubsection*{Integral relations}
\label{sssec:intrelns}

Before writing some exact and approximate solutions we first record several exact integral expressions that can be derived by directly integrating the field equations over the volume $\Bext$, being careful to keep track of its boundaries at $\rho = \rho_-$ and $\rho = \rho_+$.

Integrating the bulk Maxwell equation, \pref{BAeom}, with respect to $\rho$ in radial gauge gives
\be \label{BAeomsoln1}
  A_{\rho\theta} = A_\theta' = \frac{Q B \, e^\phi}{W^d}  \,,
\ee
for integration constant $Q$, and this in turn integrates locally to give (up to a gauge transformation)
\be \label{BAeomsoln2}
  A_\theta(\rho_+) - A_\theta(\rho_-)  = \frac{Q}{2\pi} \left \langle \frac{e^\phi}{W^{2d}} \right\rangle_{\rm ext} \,,
\ee
where we define the notation
\be \label{eq:anglebrack}
 \Bigl\langle \cdots \Bigr\rangle_{\rm ext} := \frac{1}{\sqrt{-\check g}} \int\limits_\Bext \exd^2y \, \sqrt{-g} \; \Bigl( \cdots \Bigr) = 2\pi \int\limits_{\rho_-}^{\rho_+} \exd\rho \, B W^d \, \Bigl( \cdots \Bigr) \,.
\ee

These expressions also contain the seeds of flux quantization when applied to spherical transverse dimensions. To see this we take $\rho_\pm$ to lie infinitesimally close to the north and south poles and excise these two points, leaving the topology of a sphere with two points removed (or an annulus). Integrating \pref{BAeomsoln2} around the axial direction and using the quantization of any gauge transformation, $g^{-1} \partial_\theta g$, then implies
\be \label{Fluxqndef}
 \frac{2\pi N}{g_\ssA} = \oint\limits_{\rho_-} A_\theta \, \exd \theta - \oint\limits_{\rho_+} A_\theta \, \exd \theta + Q \left \langle \frac{e^\phi}{W^{2d}} \right\rangle_{\rm ext} = \Phi_{\ssA-} + \Phi_{\ssA+} + Q \left \langle \frac{e^\phi}{W^{2d}} \right\rangle_{\rm ext} \,,
\ee
where $N$ is an integer and $g_\ssA$ is the gauge coupling for the field $A_\ssM$ and $\Phi_{\ssA\pm}$ denotes the total $A$ flux through the caps over the relevant poles. In the absence of sources at the poles we can contract the circles at $\rho = \rho_\pm$ to a point and learn $\oint_\pm A_\theta \, \exd \theta \to 0$, in which case \pref{Fluxqndef} becomes the usual condition on $Q$ required by flux quantization. But when sources are present $\oint_\pm A_\theta \, \exd \theta$ need not vanish if there is flux localized within the source. When this is so their presence on the right-hand side of \pref{Fluxqndef} modifies the implications for $Q$ of flux-quantization \cite{BLFFluxQ}.

Similarly integrating the field equation \pref{Bdilatoneom2} for the dilaton gives
\be \label{Bdilatonint}
 \Bigl[ BW^d \, \phi' \Bigr]_{\rho_+} - \Bigl[ BW^d \, \phi' \Bigr]_{\rho_-}  = \frac{\kappa^2}{2\pi} \bigl \langle V_\ssB - L_\ssA \bigr\rangle_{\rm ext} =: \frac{\kappa^2}{2\pi} \bigl\langle \cX_\ssB \bigr\rangle_{\rm ext} \,,
\ee
Two of the Einstein equations also provide direct first integrals. Integrating \pref{BXthetathetaeEinstein} leads to
\be \label{BXthetathetaeEinsteinint}
 \left[  B' W^d \right]_{\rho_+} - \left[ B' W^d \right]_{\rho_-}  = - \frac{\kappa^2}{2\pi} \left\langle \varrho_{\ssB} - \cZ_\ssB - \left( 1 - \frac{2}{d} \right) \cX_\ssB \right\rangle_{\rm ext} = - \frac{\kappa^2}{\pi} \left\langle L_{\ssA} + \frac{\cX_\ssB}{d} \right\rangle_{\rm ext} \,.
\ee
while the integral of \pref{BavR4-v1} implies
\be \label{BdwGamma=derivs}
  \left[  B \left( W^d \right)' \right]_{\rho_+} - \left[ B \left( W^d \right)' \right]_{\rho_-}  = -\frac{1}{2\pi} \left[ \check R \, \left \langle W^{-2} \right\rangle_{\rm ext} + 2\kappa^2 \bigl\langle \cX_\ssB \bigr\rangle_{\rm ext} \, \right] \,.
\ee

Of particular interest for present purposes is the simple relationship between the on-brane curvature, $\check R$, and the near-source boundary values of bulk fields obtained by combining \pref{Bdilatonint} with \pref{BdwGamma=derivs}:
\be \label{BRtildeint}
  \left[  B W^d \left( \frac{d W'}{W} + 2 \phi' \right) \right]_{\rho_+} - \left[ B W^d \left( \frac{d W'}{W} + 2 \phi' \right) \right]_{\rho_-}  = -\frac{\check R}{2\pi}  \left \langle W^{-2} \right\rangle_{\rm ext}  \,.
\ee
This states that the quantity $BW^d \left(\phi + \frac{d}{2} \, \ln W \right)'$ is monotonic in $\rho$ everywhere outside of the sources, growing or shrinking according to the sign of $\check R$ or remaining constant when $\check g_{\mu\nu}$ is flat. Should $BW^d \left(\phi + \frac{d}{2} \, \ln W \right)'$ take the same value for two different values of $\rho$ in the bulk, then we can conclude that $\check R = 0$. In particular, if there exist sources for which $2\phi' + dW'/W = 0$ at both source positions, then any interpolating geometry between the sources must satisfy $\check R = 0$; that is, the vanishing of $\left( e^\phi W^{d/2} \right)'$ in the near-source limit for both sources is a sufficient condition for $\check R = 0$ (as first argued some time ago \cite{ScaleLzero}). A special case of the $\check R = 0$ solutions are those for which $BW^d \left(\phi + \frac{d}{2} \, \ln W \right)'$ vanishes everywhere, for which $e^{\phi} \, W^{d/2}$ is constant.

\subsubsection*{Asymptotic forms}

Given the intimate relationship between $\check R$ and near-source derivatives implied by eq.~\pref{BRtildeint}, we next examine what the field equations imply about the form of the bulk solutions very close to, but outside of, a source much smaller than the transverse dimensional size ({\em eg} outside a source for which $\rho_\varv \to 0$).\footnote{Our treatment here follows closely that of \cite{Companion} and \cite{6DdS}, which are themselves based on the classic BKL treatment of near-singularity time-dependence \cite{BKL}.}

\begin{figure}[h]
\centering
\includegraphics[width=0.55\textwidth]{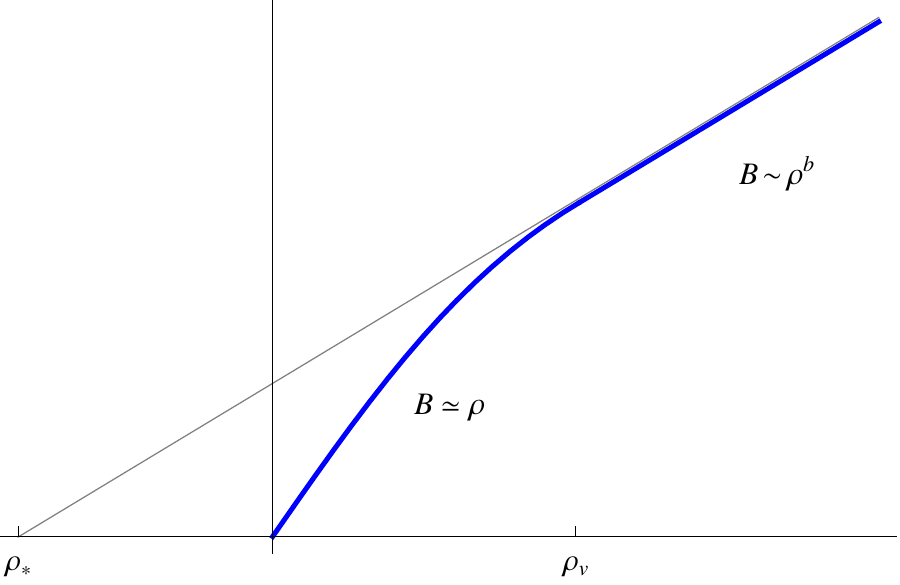}
\caption{A cartoon illustration of the definition of $\rho_\star$. The (blue) metric function $B$ increases linearly away from the origin with unit slope $B(\rho) \simeq \rho$. Outside of the source $\rho \gsim \rho_\varv$, the solution is a power law in $\rho$ with $B(\rho) \sim \rho^b$. The straight (red) line extrapolates this exterior behaviour to the point, $\rho = \rho_\star$,  where the external $B$ would have vanished if the vortex had not intervened first. }
\label{fig:rhostar}
\end{figure}

Usually the presence of such a small source induces an apparent singularity into the external geometry, at the position $\rho = \rho_\star$\footnote{Although $\rho_\star$ does not carry a source index $\varv$, it is clear that there is one such singular point for each source labelled by $\varv.$} defined as the place where $B(\rho)$ vanishes when extrapolated using only the bulk field equations. That is, it implies there would be a singularity in the external geometry if this geometry were extrapolated right down to zero size (rather than being smoothed out by the physics of the source interior, as illustrated in Figure \ref{fig:rhostar}). If the centre of the source is chosen to be $\rho = 0$ then in order of magnitude $\rho_\star$ is of order the source's size: $\rho_\star \sim \rho_\varv$.

When constructing a near-source solution it is most informative to do so as a series solution expanding in powers of the distance, $\hat\rho := \rho - \rho_\star$, from the singular source point. That is,
\bea \label{powerforms}
 W &=& W_0 \left( \frac{\hat\rho}{\ell} \right)^w + W_1 \left( \frac{\hat\rho}{\ell} \right)^{w+1} +  \cdots \nn\\
 B &=& B_0 \left( \frac{\hat\rho}{\ell} \right)^b + B_1 \left( \frac{\hat\rho}{\ell} \right)^{b+1} +  \cdots \\
 \hbox{and} \qquad
 e^\phi &=& e^{\phi_0} \left( \frac{\hat\rho}{\ell} \right)^z + \cdots \,, \nn
\eea
where $\ell$ is a measure of the proper size of the transverse geometry, which we assume to be much larger than the source's size, so $\ell \gg \hat\rho$. The powers $w$, $b$ and $z$ describe the nature of the singularity at $\rho = \rho_\star$, and all but three combinations of these parameters and the $W_i$, $B_i$ and $\phi_i$ coefficients turn out to be related to one another by the bulk field equations. The three `free' parameters are instead determined by boundary conditions that the bulk solutions satisfy in the near-source regime.

In particular, for small sources these field equations allow all of the $W_i$, $B_i$ and $\phi_i$ to be computed in terms of (say) $\phi_0$, $W_0$ and $B_0$, and further imply the following two `Kasner' relations \cite{6DdS, Kasner} amongst the powers $b$, $w$ and $z$:
\be \label{Kasnerc}
  dw + b =
 d w^2 + b^2 + z^2 = 1 \,.
\ee
The second of these in turn implies $w$, $b$ and $z$ must reside within the intervals
\be \label{limits}
 |w| \le \frac{1}{\sqrt{d}} \qquad \hbox{and} \qquad
 |b\,|, |z| \le 1 \,.
\ee
It turns out that the rest of the field equations do not give additional constraints on the three parameters $b$, $w$ and $z$, and so one combination of these is one of the quantities determined by the physical properties of the source.

\begin{figure}[h]
\centering
\includegraphics[width=\textwidth]{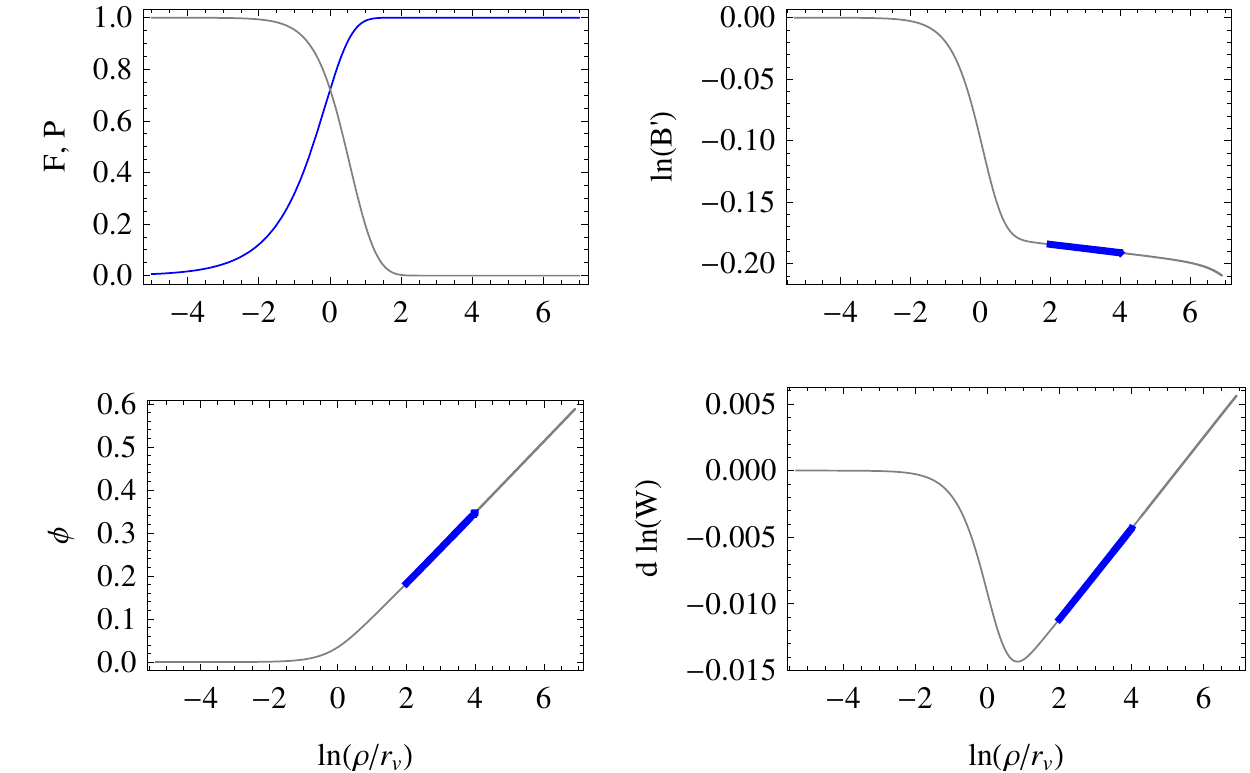}
\caption{This plot illustrates the near-source Kasner region for bulk fields that are sourced by a generic vortex with tension $v^2 = 0.25/ \kappa^2$. In the top left plot, the vortex profiles are shown on a logarithmic axis, with the (blue) scalar profile $F$ increasing towards its asymptotic vacuum value $F \to 1$ and the (grey) gauge profile $P$ decreasing towards its asymptotic value $P \to 0.$~(See below for more detailed definitions.) These profiles approach their vacuum values exponentially, and they can be neglected in the region $\tau = \ln(\rho/r_\varv) \gsim 2$, which is where the Kasner region $\rho \gsim \rho_\varv$ begins. In the next plot, $\ln (B')$ is seen to be linear in $\tau$ in the darkened (blue) near-vortex region, before the bulk sources dominate the behaviour of $B'$ at larger $\tau.$ This blue region is the Kasner region, where the bulk fields obey power laws, and this behaviour is also apparent in the plots of $\ln(e^\phi) = \phi$ and $\ln(W^d) = d \ln W.$ The Kasner powers can be extracted in this region, and for this particular source we find numerically $dw = 0.0034$, $1 - b = 0.0034$ and $z = 0.082$ which is in agreement with the perturbative solution of \protect\pref{eq:kasnersolved}. Finally, we note that these numerical results are also consistent with the estimates $z \sim \delta \sim \kappa^2 v^2 $ as is argued below.}
\label{fig:bigkasner}
\end{figure}

Expansion about a regular point --- {\em ie} not the location of a singularity --- should correspond to the specific solution $z = w = 0$ and $b = 1$ to eqs.~\pref{Kasnerc}. In the presence of weakly gravitating sources we expect to find small deviations from these values, whose size can be inferred by solving \pref{Kasnerc} perturbatively. To this end we write
$b = 1 + b_1 \delta + b_2 \delta^2 + \cdots$, $w = w_1 \delta + \cdots$ and $z = z_1 \delta + \cdots$ and expand. Working to order $\delta^2$ we find the one-parameter family of solutions
\be \label{eq:kasnersolved}
 z = z_1 \delta + \cO(\delta^2)\,, \qquad
 b = 1-\frac{z^2}{2} + \cO(\delta^3)\qquad \hbox{and} \qquad
 dw = \frac{z^2}{2} + \cO(\delta^3)\,.
\ee
Quite generally only $z$ can deviate from its background value at linear order, and the leading quadratic contributions to $w$ and $b-1$ are not independent of this linear deviation in $z$. In later sections we find $z$ is determined by the coupling of $\phi$ to the source lagrangian, and so we generically expect $\delta$ to be of order $\kappa^2 v^2$, where $v^2$ is a measure of the energy density (or tension) carried by the source (more about which below). Fig.~\ref{fig:bigkasner} provides numerical evidence for the validity of the Kasner equations and their perturbative solution, \pref{eq:kasnersolved}, nearby but outside of an explicit vortex solution.

In terms of these Kasner exponents, the combination appearing on the left-hand side of eq.~\pref{BRtildeint} in the near-source limit is
\be \label{LeadingKasR}
   \lim_{\rho \to \rho_\star} B W^d \left( \frac{d W'}{W} + 2 \phi' \right) = (2z + dw) \left( \frac{B_0 W_0^d}{\ell} \right) \left( \frac{\hat\rho}{\ell} \right)^{dw+b-1} = (2z + dw) \left( \frac{B_0 W_0^d}{\ell} \right) \,,
\ee
where the second equality uses the linear Kasner condition \pref{Kasnerc}. When eq.~\pref{LeadingKasR} is order unity then eq.~\pref{BRtildeint} implies that the curvature $\check R$ is of order $\check R \sim (\kappa_d^2/\kappa^2)$ corresponding to a $d$-dimensional energy density of order $1/\kappa^2$. Here $\kappa_d$ is the dimensionally reduced gravitational coupling of the low-energy $d$-dimensional theory, and we use a result, derived below, that $\kappa_d^{-2} \simeq \kappa^{-2} \langle W^{-2} \rangle_{\rm ext}$.

Of particular interest are situations where the near-source solutions satisfy $z = 0$ since the Kasner conditions imply these also satisfy $w = 0$ and so $2z + dw = 0$. Consequently for any such source the leading contribution, eq.~\pref{LeadingKasR}, to $\check R$ vanishes. In this case it is a subleading term that first contributes on the left-hand side of \pref{BRtildeint}, implying that $\check R$ is suppressed relative to $1/\kappa^2$ by a power of the ratio of source and extra-dimensional sizes: $\rho_\pm/\ell $. This asymptotic reasoning is borne out by the explicit numerical and analytic solutions described in the next sections.

\subsubsection*{Explicit bulk solutions}

It is useful to see how the above general arguments go through for explicit solutions to the bulk field equations, which are known in great detail \cite{GGP, DistributedSUSY} when $\phi' \to 0$ in the near-source limit in the special case where $d = 4$ and $D = 6$ and $V_0 = 2 g_\ssR^2/\kappa^4$. (See also \cite{Linear6DSugra} for a discussion of linearized solutions; \cite{6DdS, 6DdS2} for solutions with\footnote{The de Sitter solutions to these equations are interesting in their own right as a counter-examples \cite{6Dnogonot} to no-go theorems for the existence of de Sitter solutions in supergravity \cite{6Dnogo}.} $\phi' \ne 0$; \cite{Swirl} with other fields nonzero; \cite{TimeDep} for time-dependent configurations; \cite{MultiBrane} with more than two brane sources; and with black hole solutions \cite{Nemanja6DBH}.)

The solutions are most simply written using the symmetry ansatz
\be \label{eq:GGPmetric}
 \exd s^2 = W^2(\xi) \, \exd s_4^2 + r^2(\xi) \Big(\exd\xi^2 + \alpha^2(\xi) \sin^2\xi \,\exd\theta^2 \Big) \,,
\ee
where $\exd s_4^2$ denotes the maximally symmetric on-vortex geometry, $\exd s_4^2 = \check g_{\mu\nu} \,\exd x^\mu \exd x^\nu$. With this ansatz, as seen above, the field equations ensure $\check R = 0$ provided we assume $\phi' \to 0$ in the near-source limit, which we now do (and so also take $\check g_{\mu\nu} = \eta_{\mu\nu}$). The dilaton and metric function then turn out to be
\be
 e^{\phi(\xi)} = \frac{e^{\varphi}}{W^2(\xi)}\,, \qquad
 \alpha(\xi) = \frac\Upsilon{W^4(\xi)} \qquad \hbox{and} \qquad r(\xi) =  r_\ssB W(\xi) e^{-\varphi/2} \,,
\ee
with
\be
 W^4(\xi)  = e^{\upsilon} \sin^2\frac{\xi}{2} + e^{-\upsilon} \cos^2 \frac{\xi}2
 = \cosh \upsilon - \sinh \upsilon \,\cos\xi \,,
\ee
where $\upsilon$, $\Upsilon$ and $\varphi$ are integration constants. Notice that $r^2 e^{\phi} = r_\ssB^2$ for all $\xi$, with the length-scale $r_\ssB$ set by
\be
 r_\ssB := \frac{\kappa}{2\gR} \,.
\ee

The two sources for this geometry are located at the two singular points, $\xi_- = 0$ and $\xi_+ = \pi$, where the transverse space pinches off. In the near-brane limit, $\xi \to 0$, we have $W \to W_- + \cO(\xi^2)$, and so the proper radial distance, given by
\be
 \rho(\xi) = \int \exd \xi \; r(\xi) = r_\ssB \, e^{-\varphi/2} \int \exd \xi \; W(\xi) \,,
\ee
in this limit becomes $\rho = r_\ssB \, e^{-\varphi/2} \left[ W_0 \xi + \cO( \xi^3) \right]$, and so $\alpha \to \alpha_- = \Upsilon/W_-^4 + \cO(\rho^2)$, $B \to \alpha_- \rho + \cO(\rho^3)$ and $e^\phi \to e^\varphi/W_-^2 + \cO(\rho^2)$ in the near-brane limit (corresponding to the Kasner exponents $z = w = 0$ and $b = 1$). Similar properties also hold for $\xi \to \pi$.

Two of the integration constants --- $\upsilon$ and $\Upsilon$ --- can be traded for the conical defect angles, $\delta_\pm = 2\pi(1 - \alpha_\pm)$, in the two near-brane limits, with
\be \label{eq:Wpmforms}
 \Upsilon = \sqrt{\alpha_+\alpha_-} \qquad \hbox{and} \qquad e^{\upsilon} = \sqrt{\frac{\alpha_-}{\alpha_+}} = W_+^4 = \frac{1}{W_-^4} \,.
\ee
In terms of these $\alpha(\xi)$ and $W(\xi)$ are given simply by
\be
 \frac1{\alpha(\xi)} = \frac{1}{\alpha_-} \; \cos^2 \frac{\xi}2 + \frac{1}{\alpha_+} \; \sin^2\frac{\xi}{2} \quad \hbox{and} \quad
 W^4(\xi) = W_-^4 \cos^2 \frac{\xi}2 + W_+^4 \sin^2\frac{\xi}{2} \,.
\ee
In particular, in the special case $W_+ = W_-$ the function $W(\xi)$ --- and so also $\phi(\xi)$, $r(\xi)$ and $\alpha(\xi)$ --- becomes constant, and the geometry reduces to the simple rugby-ball solution \cite{SLED}. This has proper distance $\rho = \xi \, r_\ssB \, e^{-\varphi/2}$ and
\be
 B(\rho) = \alpha \, r_\ssB \, e^{-\varphi/2} \, \sin \left( \frac{\rho}{r_\ssB e^{-\varphi/2}} \right) =: \alpha \, \ell \, \sin \left( \frac{\rho}{\ell} \right) \,,
\ee
and so $\ell = r_\ssB \, e^{-\varphi/2} = \frac12 \kappa/g_\ssR(\varphi)$ --- where $g_\ssR(\varphi) = g_\ssR \, e^{\varphi/2}$ --- represents the proper `radius' of the transverse dimensions, whose physical volume is $\Omega = 4 \pi \ell^2$.

The gauge field for the general solution with different brane properties is given by
\be
 A_{\xi\theta} =  \frac{\cQ \, \Upsilon \,\sin\xi}{2 g_\ssA \, W^8(\xi)}
  \,,
\ee
where $g_\ssA$ is the corresponding gauge coupling constant. Comparing with \pref{BAeomsoln1} shows $\cQ$ is related to $Q$ by $Q = \cQ /(2 g_\ssA r_\ssB^2)$ and so is not an independent constant. The total amount of flux present in the bulk (region $\Bext$) then is
\be
 \int A_{(2)} = \int_-^+ \!\exd\xi \exd \theta \; A_{\xi\theta} = \frac{2\pi \cQ \Upsilon}{g_\ssA} \,,
\ee
so flux quantization, \pref{Fluxqndef}, relates any source flux to the defect angles by
\be \label{Fluxqnappl}
 N = \frac{g_\ssA \Phi_{\ssA-}}{2\pi} + \frac{g_\ssA \Phi_{\ssA+}}{2\pi} + \cQ \Upsilon \,.
\ee
This shows why brane-localized flux is generically necessary to satisfy flux quantization for two branes with generic tensions if $\cQ$ is otherwise fixed. When there is no localized flux, we can estimate
\be
Q \sim \frac{ 1}{g_\ssA r_\ssB^2} \sim \frac{ g_\ssR^2 }{g_\ssA \kappa^2} \sim \frac{ g_\ssR }{ \kappa^2 }\,.
\ee

\section{Sources - the UV picture}
\label{sec:sourcesystem}

The previous section describes several exact consequences of the bulk field equations that relate bulk properties to the asymptotic near-source behaviour of various combinations of bulk fields. In particular it shows how the on-source curvature, $\check R$, is determined in this way purely by the near-source combination $BW^d \left(\phi + \frac{d}2 \, \ln W \right)'$, and so vanishes in particular when $\phi'$ and $W'$ approach zero in this limit.

This section now turns to the question of how these derivatives are related to source properties, extending the results of \cite{Companion} to include dilaton couplings and extending those of \cite{OtherGVs} to include nonzero brane-localized flux (more about which below). In this section this is done by making an explicit construction of the sources within a UV completion, as a generalization of Nielsen-Olesen vortices \cite{NOSolns,CStrings}. We do so by adding new scalar and gauge fields that admit such vortex solutions, with a view to understanding in more detail how near-source behaviour is controlled by the source properties. Because our focus here is mostly on classical issues we do not explicitly embed the new sector into a supersymmetric framework, but we return to this issue when considering quantum corrections in subsequent analysis \cite{STMicro}.

A key assumption in our discussion is that the typical transverse vortex size, $\hat r_\varv$, is much smaller than the size, $\ell$, of the transverse external space: $\hat r_\varv \ll \ell$. Subsequent sections then re-interpret the results found here in terms of the $D$-dimensional IR effective theory applying over length scales $\hat r_\varv \ll \rho \lsim \ell$. We follow closely the discussion of \cite{Companion}, highlighting the differences that arise as we proceed (the main one of which is the presence of the dilaton zero mode).

\subsection{Action and field equations}
\label{subsec:actionFE2}

We start with the action and field equations for the UV completed system describing the sources. With Nielsen-Olesen solutions in mind, we take this `vortex' --- or `brane' --- sector to consist of an additional complex scalar coupled to a second $U(1)$ gauge field. Again we work in $D = d+2$ spacetime dimensions, with the cases $(D,d) = (4,2)$ and $(D,d) = (6,4)$ being of most later interest.

The full action now is $S = S_\ssB + S_\ssV$ with $S_\ssB$ as given in \pref{SB} and the vortex part of the action given explicitly by
\bea \label{SV}
 S_\ssV &=& - \int \exd^{d+2}x \; \sqrt{-g} \left[ \frac14  \,e^{p \phi} Z_{\ssM \ssN}^2 + \frac{\varepsilon}2  \,e^{r \phi} Z_{\ssM \ssN} A^{\ssM \ssN} + D_\ssM \Psi^* \, D^\ssM \Psi  + \lambda \, \,e^{q \phi} \left(\Psi^* \Psi - \frac{v^2}{2} \right)^2 \right] \nn\\
 &=:& - \int \exd^{d+2}x \; \sqrt{-g} \; \Bigl( L_\ssZ + L_{\rm mix} + L_\Psi + V_b \Bigr) =: - \int \exd^{d+2}x \; \sqrt{-g} \;  L_\ssV\,,
\eea
where $D_\ssM \Psi := \partial_\ssM \Psi - i e Z_\ssM \, \Psi$, and the second line defines the various $L_i$. The terms $L_\ssZ$, $L_\Psi$ and $V_b$ describe scalar electrodynamics and are chosen to allow vortex solutions for which the $Z_{\ssM\ssN}$ gauge flux is localized. The $L_{\rm mix}$ term is chosen --- following \cite{Companion} (see also \cite{GS, DStrings}) --- to kinetically mix $Z$ with the bulk gauge field \cite{Bob} and thereby generate a vortex-localized component to the exterior $A_{\ssM\ssN}$ gauge flux.

We follow \cite{Companion} and write $\sqrt{2} \; \Psi = \psi \, e^{i \Omega}$ and adopt a unitary gauge for which the phase, $\Omega$, is set to zero, though this gauge will prove to be singular at the origin of the vortex solutions we examine later. In this gauge the term $L_\Psi$ in $S_\ssV$ can be written
\be
 L_\Psi = D_\ssM \Psi^* D^\ssM \Psi = \frac12 \Bigl( \partial_\ssM \psi \, \partial^\ssM \psi + e^2 \psi^2 Z_\ssM Z^\ssM \Bigr)
\ee
and the vortex potential becomes
\be
 V_b(\phi,\psi) = \frac{\lambda}4 e^{q\phi} \Bigl( \psi^2 - v^2 \Bigr)^2 \,.
\ee
Our interest in what follows is largely in how the dependence on $\phi$ in these interactions back-reacts onto the properties of the bulk, and affects the interactions of the various low-energy effective descriptions. We notice in passing that $S_\ssV$ shares the classical scale invariance of $S_\ssB$ only in the special case: $p=r=-1$ and $q=1$.

It is also useful to group the terms in the vortex and bulk lagrangians together according to how many metric factors and derivatives appear, with
\bea \nn
 &&\phantom{OO}L_{\rm kin} :=  \frac12 \, g^{\ssM\ssN} \left( \frac{1}{\kappa^2} \, \partial_\ssM \phi \, \partial_\ssN \phi + \partial_\ssM \psi \, \partial_\ssN \psi \right)\,, \qquad
 L_{\rm gge} := L_\ssA + L_\ssZ + L_{\rm mix} \\
 &&\qquad\qquad \qquad L_{\rm pot} := V_\ssB(\phi) + V_b(\phi, \psi) \qquad
 \hbox{and} \qquad
 L_{\rm gm} := \frac12 \,e^2 \psi^2 \, g^{\ssM \ssN} Z_\ssM Z_\ssN \,,
\eea
so $L_\phi + L_\Psi + V_b = L_{\rm kin} + L_{\rm pot} + L_{\rm gm}$. Notice that for the configurations of later interest we have $L_\ssZ \ge 0$, $L_\Psi \ge 0$ and $V_b \ge 0$ while $L_{\rm mix}$ can have either sign.

\subsubsection*{Gauge field equations}

With these choices the field equations for the two Maxwell fields are
\be \label{checkAeq}
 \frac{1}{\sqrt{-g}} \, \partial_\ssM \Bigl[ \sqrt{-g} \Bigl( e^{-\phi} A^{\ssM \ssN} + \varepsilon \, e^{r\phi} Z^{\ssM \ssN} \Bigr) \Bigr] = 0 \,,
\ee
and
\be \label{Z0eq}
 \frac{1}{\sqrt{-g}} \, \partial_\ssM \Bigl[ \sqrt{-g} \Bigl( e^{p\phi} Z^{\ssM \ssN} + \varepsilon \, e^{r\phi}  A^{\ssM \ssN} \Bigr) \Bigr] = e^2 \psi^2 Z^\ssN  \,,
\ee
and (as usual) these are supplemented by the Bianchi identities $\exd A = \exd Z = 0$, for the 2-forms $A_{\ssM\ssN}$ and $Z_{\ssM\ssN}$. For later purposes it is useful to write \pref{checkAeq} as $\partial_\ssM \Bigl( \sqrt{-g} \; e^{-\phi} \check A^{\ssM\ssN} \Bigr) = 0$ with $\check A_{\ssM\ssN}$ defined by
\be \label{AZcheckA}
  \check A_{\ssM\ssN} :=  A_{\ssM\ssN} + \varepsilon \, e^{(r+1)\phi} Z_{\ssM\ssN} \,,
\ee
in which case \pref{Z0eq} becomes
\be \label{Zcheckeq}
 \frac{1}{\sqrt{-g}} \, \partial_\ssM \Bigl[ \sqrt{-g} \; \Lambda(\phi) \,  Z^{\ssM \ssN} \Bigr] + \varepsilon (r+1) \, e^{r\phi} \partial_\ssM \phi
 \, \check A^{\ssM \ssN} = e^2 \psi^2 Z^\ssN  \,,
\ee
with
\be \label{Lambdadef}
 \Lambda(\phi) := e^{p\phi} - \varepsilon^2 \, e^{(2r+1)\phi} \,.
\ee

We see below that the energy density of the system is given by a particular sum of the $L_i$'s, and when assessing the sign of the energy it is useful to notice that the off-diagonal contribution to $L_{\rm gge}$ vanishes when this is expressed in terms of $\check A_{\ssM\ssN}$ rather than $A_{\ssM\ssN}$, since
\be
 L_{\rm gge} = L_\ssA + L_\ssZ + L_{\rm mix} = \check L_\ssA + \check L_\ssZ \,,
\ee
where
\be
 \check L_\ssA := \frac14 e^{-\phi} \check A_{\ssM \ssN} \check A^{\ssM \ssN}
 \quad \hbox{and} \quad
 \check L_\ssZ := \frac14 \, \Lambda(\phi) \,  Z_{\ssM \ssN}Z^{\ssM \ssN} \,.
\ee
This shows that the kinetic energy of the $Z_\ssM$ gauge field is renormalized\footnote{This provides a UV perspective to what becomes a divergent renormalization \cite{ 6DSUSYUVCaps, UVCaps,ClassRenorm} in the limit of zero-size sources.} by the mixing of the two gauge fields, with the result only bounded below for all\footnote{For some applications requiring boundedness for all $\phi$ may be too strong a criterion, since the semiclassical approximation relies on the assumption $e^\phi \ll 1$. Since our inference about the boundedness of the energy is itself performed semiclassically, it is also suspect if the unboundedness occurs only when $e^\phi \gsim 1$. If this weaker criterion is our guide then we only require $2r+1 \ge p$ rather than strict equality.} real $\phi$ if $p = 2r+1$ and $\varepsilon^2 < 1$. It also suggests a better split between the bulk and the vortices is to write $L_\ssB + L_\ssV = \check L_\ssB + L_{\rm loc}$, where $\check L_\ssB = L_\EH + L_\phi + \check L_\ssA$ and $L_{\rm loc} = \check L_\ssZ + L_\Psi + V_b$. Split this way all of the localized energy falls within $L_{\rm loc}$, because of the absence of mixing terms \cite{Companion}.

Although the quantities $\check L_\ssA$ and $\check L_\ssZ$ are useful when describing the energy density, unlike in \cite{Companion} their use directly in the lagrangian can lead to errors. It is important in this regard to keep in mind that there are two important ways in which the transition from $A_{\ssM\ssN}$ to $\check A_{\ssM\ssN}$ differs in the present case from the discussion in \cite{Companion}. First, if $\check A_{\ssM\ssN}$ is used instead of $A_{\ssM\ssN}$ in the field equations then one must remember that the Bianchi identity for $A$ and $Z$ implies $\check A$ also satisfies
\be \label{newBianchi}
 \exd \check A = \varepsilon (r+1) \, e^{(r+1)\phi} \; \exd \phi \wedge Z \,,
\ee
which need not vanish because of the presence in \pref{AZcheckA} of the field $\phi$. The second difference is related to the first: one must {\em never} use eq.~\pref{AZcheckA} to trade $A_{\ssM\ssN}$ for $\check A_{\ssM\ssN}$ in the action and {\em then} compute the field equations. This is because \pref{AZcheckA} is not a change of variables in the path integral since it is not a redefinition of the gauge potentials. In practice this kind of substitution is most dangerous in the $\phi$ field equation, as may be seen from the functional chain rule,
\bea
 \left( \frac{\delta S}{\delta \phi(x)} \right)_{A \, {\rm fixed}} &=&
 \left( \frac{\delta S}{\delta \phi(x)} \right)_{\check A \, {\rm fixed}} + \int \exd^Dy \;  \left( \frac{\delta S}{\delta \check A_{\ssM\ssN}(y)}  \right)_{\phi \, {\rm fixed}}  \left( \frac{\delta \check A_{\ssM\ssN}(y)}{\delta \phi(x)} \right)_{A \, {\rm fixed}} \nn\\
 &=& \left( \frac{\delta S}{\delta \phi(x)} \right)_{\check A \, {\rm fixed}} + \varepsilon (r+1) \, e^{(r+1)\phi} Z_{\ssM\ssN} \left( \frac{\delta S}{\delta \check A_{\ssM\ssN}(x)}  \right)_{\phi \, {\rm fixed}} \,.
\eea
The second term on the right-hand side of this relation need {\em not} vanish when the field equations are satisfied.

For configurations with the symmetries of interest the gauge field equations reduce to
\be \label{Acheckeom}
  \left( \frac{ e^{-\phi} W^d  \check A_{\rho\theta}}{B} \right)' = 0 \,,
\ee
and
\be \label{Zeom}
 \frac{1}{BW^d} \, \left[ \Lambda(\phi) \left( \frac{ W^d Z_\theta'}{B} \right) \right]' + \varepsilon (r+1) \, e^{r\phi} \, \left( \frac{\phi' \check A_{\rho\theta}}{B^2} \right) = \frac{e^2 \psi^2 Z_\theta}{B^2}  \,,
\ee
where (as before) primes denote differentiation with respect to proper distance, $\rho$, and $\Lambda(\phi)$ and $\check A_{\ssM\ssN}$ are as defined in eqs.~\pref{Lambdadef} and \pref{AZcheckA}, respectively.

\subsubsection*{Flux quantization}

The solution to eq.~\pref{Acheckeom} is
\be
 \check A_{\rho\theta} = \frac{Q \, B\, e^\phi}{W^d} \,,
\ee
which shows that $\check A_{(2)}$ describes the part of the gauge fields that does not see the vortex sources. Ultimately the integration constant $Q$ is fixed by the flux quantization conditions for $A_{(2)}$ and $Z_{(2)}$, which state
\be
 \Phi_\ssA := \int A_{(2)} = \int \exd^2y \; A_{\rho\theta} = \frac{2\pi N}{g_\ssA} \,,
\ee
and
\be
 \Phi_\ssZ := \int Z_{(2)} = \int \exd^2y \; Z_{\rho\theta} = -\sum_\varv \frac{2\pi n_\varv}{e} = -\frac{2\pi n_{\rm tot}}{e}
 \,,
\ee
where $N$ and $n_{\rm tot} = \sum_\varv n_\varv = n_+ + n_-$ are integers while $e$ and $g_\ssA$ are the relevant gauge couplings. Strictly speaking flux quantization only ensures the sum over all vortices, $n_{\rm tot} = n_+ + n_-$, is an integer. However we imagine here that the two vortices are situated at opposite ends of the (relatively) very large extra dimensions and so are very well-separated. Consequently in practice each of $n_+$ and $n_-$ are separately integers, up to exponential accuracy.\footnote{This could also be alternatively arranged by having two copies of the vortex sector, with each vortex carrying flux from a different $U(1)$ (which would therefore be separately quantized) but with both mixing with the bulk $U(1)$.}

On one hand, for $\phi \approx \phi_\varv$ approximately constant across the narrow width of each vortex, this implies
\be
 \check \Phi_\ssA := \int \check A_{(2)} \approx 2\pi \left( \frac{N}{g_\ssA} - \frac{ \varepsilon }{e} \, \sum_\varv n_\varv e^{(r+1)\phi_\varv}    \right)
 \,,
\ee
while on the other hand the left-hand side is related to $Q$ by
\be
  \check \Phi_\ssA = Q \int \exd^2y \;  \left( \frac{B \, e^\phi}{W^4} \right) = Q\, \widehat \Omega_{-4} \,,
\ee
where we define the useful notation
\be \label{Omegakdef}
 \widehat \Omega_k  := \int \exd^2y \, \sqrt{g_2} \, W^{k}\,  e^\phi  = \int \d^2 y \, \sqrt{\hat g_2} W^k \,,
\ee
to represent the 2D integrals that arise here and in later calculations. Here $\widehat \Omega_k$ is the integral of $W^k$ over the transverse dimensions using the scale-invariant metric, $\hat g_{mn} := e^\phi \, g_{mn}$, and the particular case $k = 0$ represents the extra-dimensional volume, $\widehat \Omega := \widehat \Omega_0$, as measured by this metric.

We see $Q$ is given by
\be \label{fluxqn}
 Q = \frac{\check \Phi_\ssA}{\widehat \Omega_{-4}} \approx \frac{2\pi}{\widehat \Omega_{-4}} \left( \frac{N}{g_\ssA} - \frac{\varepsilon}{e} \sum_\varv n_\varv \, e^{(r+1)\phi_\varv}  \right)
 \,,
\ee
In the special case where $\phi = \varphi$ takes the same value at all of the vortex positions this becomes
\be \label{fluxqnvar}
 Q \approx \frac{2\pi}{\widehat\Omega_{-4} } \left[ \frac{N}{g_\ssA} - \left( \frac{\varepsilon \, n_{\rm tot}}{e} \right)  e^{(r+1)\varphi}  \right]
 \,.
\ee

\subsubsection*{Scalar field equations}

The vortex scalar field equation in unitary gauge becomes
\be \label{Psieom}
 \frac{1}{\sqrt{-g}} \, \partial_\ssM \Bigl( \sqrt{-g} \; g^{\ssM \ssN} \partial_\ssN \psi \Bigr) = e^2 \psi Z_\ssM Z^\ssM + \lambda \,e^{q\phi} \psi \Bigl(\psi^2 - v^2 \Bigr) \,,
\ee
while the dilaton equation is
\bea \label{dilatoneq}
 \Box \phi = \frac{1}{\sqrt{-g}} \, \partial_\ssM \Bigl( \sqrt{-g} \; g^{\ssM\ssN} \partial_\ssN \phi \Bigr) &=& \kappa^2 \Bigl( V_\ssB - L_\ssA + q V_b + p L_\ssZ + r L_{\rm mix} \Bigr) \\
 &=& \kappa^2 \Bigl( \cX + \cY \Bigr) \,. \nn
\eea
Here
\be \label{Xdef}
 \cX := L_{\rm pot} - L_{\rm gge} = V_\ssB + V_b - L_\ssA - L_\ssZ - L_{\rm mix} \,,
\ee
is the combination appearing in the stress tensor, ${T^m}_m = -2 \cX$, and we define the useful quantity
\be \label{Ydef}
 \cY := (q-1) V_b + (1+p) L_\ssZ + (1+r) L_{\rm mix} \,.
\ee
Notice that $\cY$ involves only terms from the vortex lagrangian and vanishes identically in the scale-invariant case, for which $p = r = -1$ and $q = 1$, while $\cX$ is most usefully split between bulk and localized contributions through $\cX = \check \cX_\ssB + \cX_{\rm loc}$ where $\check\cX_\ssB = V_\ssB - \check L_\ssA$ while $\cX_{\rm loc} := V_b  - \check L_\ssZ$. It is similarly useful to write $\cZ = \cZ_\ssB + \cZ_{\rm loc} $ with $\cZ_\ssB$ defined as above and $\cZ_{\rm loc} = \frac{1}{2} (\psi^\prime)^2 - L_{\rm gm}.$

Once restricted to the symmetric configurations of interest the scalar equations simplify to
\be \label{Psieom2}
 \frac{1}{BW^d} \, \Bigl( BW^d \, \psi' \Bigr)' = e^2 \psi \left( \frac{Z_\theta}{B} \right)^2 + \lambda e^{q\phi} \psi \Bigl(\psi^2 - v^2 \Bigr) \,,
\ee
and
\be \label{dilatoneom2}
 \frac{1}{BW^d} \, \Bigl( BW^d \, \phi' \Bigr)' = \kappa^2 \Bigl(\cX + \cY \Bigr) \,.
\ee

\subsubsection*{Einstein equations}

The stress-energy tensor of the entire system now is
\bea \nn
 T_{\ssM\ssN} &=&  \frac{1}{\kappa^2} \, \partial_\ssM \phi \, \partial_\ssN \phi +  \partial_\ssM \psi \, \partial_\ssN \psi + e^2 \psi^2 Z_\ssM Z_\ssN + e^{-\phi} A_{\ssM\ssP} {A_\ssN}^\ssP + e^{p\phi} Z_{\ssM \ssP} {Z_\ssN}^\ssP \\
  && \qquad + \frac{\varepsilon}2 e^{r\phi} \Bigl( A_{\ssM\ssP} {Z_\ssN}^\ssP + Z_{\ssM \ssP} {A_\ssN}^\ssP \Bigr) - g_{\ssM \ssN} \Bigl( L_{\rm kin} + L_{\rm gm} + L_{\rm pot} + L_{\rm gge} \Bigr) \,,
\eea
and so the nontrivial components of the matter stress-energy are given by \pref{Tmunusymform}, with $\cX = L_{\rm pot} - L_{\rm gge}$ (as in eq.~\pref{Xdef}) while
\be
 \varrho :=  L_{\rm kin} + L_{\rm gm} + L_{\rm pot} + L_{\rm gge} \qquad \hbox{and} \qquad
 \cZ := L_{\rm kin} - L_{\rm gm} \,.
\ee

The lone nontrivial component of the trace-reversed Einstein equations governing the $d$-dimensional on-brane geometry therefore becomes
\be \label{avR4-v1}
 \cR_{(d)} := g^{\mu\nu} \cR_{\mu\nu} = \frac{\check R}{W^2} + \frac{d}{BW^d} \Bigl( BW' W^{d-1} \Bigr)'
 = -2\kappa^2 \cX \,.
\ee
The two components dictating the 2-dimensional transverse geometry similarly can be taken to be
\be \label{avR2}
 \cR_{(2)} := g^{mn} \cR_{mn} = R + d \left( \frac{W''}{W} + \frac{B'W'}{BW} \right) = - \kappa^2 {X^m}_m = - 2 \kappa^2 \left[  \varrho - \left( 1 - \frac{2}{d} \right) \cX \right] \,,
\ee
and
\be \label{newEinstein}
 \frac{B}{W} \left( \frac{W'}{B} \right)' = - \frac{2}{d} \; \kappa^2 \cZ \,.
\ee
For later purposes a useful combination of these equations gives the $D$-dimensional Ricci scalar, $\cR_{(D)} = \cR_{(d)} + \cR_{(2)}$. Given that the total lagrangian is given by $L = L_\ssB + L_\ssV = (2\kappa^2)^{-1} \cR_{(D)} + \varrho $, where $\varrho$ is the total energy density, it turns out this can be written
\be \label{LtotvsX}
 L = L_\ssB + L_\ssV = - \frac{2\cX}{d}
\ee

Yet another useful combination of the above equations is the $(\rho \rho)$ component of the Einstein equations $\cG^\rho{}_\rho = - \kappa^2 T^{\rho}{}_\rho$ which reads
\be \label{hamconstraint}
 2 d \left( \frac{B' W'}{BW} \right) + \frac{\check R}{W^2} + d(d-1) \left( \frac{W'}{W} \right)^2 = 2\kappa^2 \left( \cZ  - \cX \right) \,.
\ee
This expression contains only first derivatives of metric and matter fields, and acts as a constraint on the solution as it is integrated along the proper distance $\rho.$

Finally, we see from the above that the gauge fields only enter the Einstein equations through the combination $L_{\rm gge} = L_\ssA + L_\ssZ + L_{\rm mix} = \check L_\ssA + \check L_\ssZ$, and so the Einstein equations are indifferent (unlike the dilaton field equation) to whether they are expressed using $A_{\ssM\ssN}$ or $\check A_{\ssM\ssN}$.

\subsubsection*{Control of approximations}
\label{subsec:scales}

Since solutions to the classical field equations take up much of what follows, we first briefly digress to summarize the domain of validity of these solutions. The fundamental parameters of the problem are the gravitational constant, $\kappa$; the coefficient of the bulk scalar potential, $V_0$ (or $V_0 = 2g_\ssR^2/\kappa^4$ for 6D supergravity); the gauge couplings, $\hat e^2(\varphi) = e^2/\Lambda(\varphi)$ and $\hat g_\ssA(\varphi) = g_\ssA e^{\varphi/2}$; the scalar self-coupling, $\hat\lambda(\varphi) = \lambda e^{q\varphi}$, and the scalar vev $v$. To these must be added the dimensionless parameter, $\varepsilon$, that measures the mixing strength for the two gauge fields. When discussing 6D supergravity we typically assume $g_\ssA = g_\ssR$ and so can use $g_\ssA$ and $g_\ssR$ interchangeably. We also largely keep to the vortex parameter range $\hat\lambda \sim \hat e^2$.

The energy density of the vortex turns out below to be of order $\hat e^2 v^4$ and when $\hat \lambda \sim \hat e^2$ the transverse vortex proper radius is of order $\hat r_\varv$ with
\be
 \hat r_\varv = \frac{1}{\hat e v} \,.
\ee
The effective energy-per-unit-area of the vortex is therefore of order $\hat e^2 v^4 \hat r_\varv^2 = v^2$. These energies give rise to $D$-dimensional curvatures within the vortex of order $1/l_\varv^2 = \kappa^2 \hat e^2 v^4$ and integrated dimensional gravitational effects (like conical defect angles) of order $\kappa^2 v^2$. We work in a regime where $\kappa v \ll 1$ to ensure that the gravitational response to the energy density of the vortex is weak, and so defect angles are small and $l_\varv \gg \hat r_\varv$. We also define the $\phi$-independent quantity $r_\varv = 1/e v.$

By contrast, we have seen that far from the vortex the curvature scale in the bulk turns out to be proportional to $1/\ell^2$ where
\be
 \ell = r_\ssB \, e^{-\varphi/2} = \frac{\kappa}{2 \hat g_\ssR(\varphi)} \,.
\ee
Since our interest is in the regime where the vortex is much smaller than the transverse dimensions we throughout assume $\hat r_\varv / \ell \ll 1$ and so the parameter range of interest is
\be
  \frac{\hat g_\ssA(\varphi)}{\hat e(\varphi)}  \ll \kappa v \ll 1 \,.
\ee

As seen earlier, semiclassical reasoning also depends on the ambient value of the dilaton, $\varphi$, because it is $e^{d\varphi/2}$ that counts loops in the bulk theory. Consequently we require
\be
 e^{\varphi} \ll 1
\ee
in order to work semiclassically within the bulk theory. But $\varphi$ also governs the size of vortex couplings through $\hat \lambda(\varphi) = \lambda e^{q \varphi} $ and $\hat e^2(\varphi) = e^2 / \Lambda(\varphi)$ and we must check these remain small to trust semiclassical reasoning on the vortex.

\subsection{Dual formulation}
\label{sec:dualform}

Because the gauge coupling to the vortex is magnetic, it can be useful to work with the Hodge dual of the Maxwell field $A_{\ssM\ssN}$. In this section we restrict to the case of later interest, $(D,d) = (6,4)$, though the same steps are easily also done for general $d$.

The terms involving $A_\ssM$ in the 6D action can be written
\be
 L_\ssA + L_{\rm mix} + L_{\rm lm} = \frac14 \, e^{-\phi} A_{\ssM \ssN} A^{\ssM \ssN} + \frac{\varepsilon}2 \, e^{r\phi} Z^{\ssM\ssN} A_{\ssM \ssN} + \frac{1}{3!} \, \epsilon^{\ssM\ssN\ssP\ssQ \ssR\, \ssT} V_{\ssM\ssN\ssP} \partial_\ssQ A_{\ssR\ssT} \,,
\ee
where the functional integration over the 3-form lagrange multiplier, $V_{\ssM\ssN\ssP}$, ensures $A_{\ssM\ssN}$ satisfies the Bianchi identity and so allows us to directly integrate $A_{\ssM\ssN}$ rather than the gauge potential, $A_\ssM$. Notice that because we wish the constraint $\exd A_{(2)} = 0$ also to hold on any boundaries we do not include a surface term to restrict the variation of $V_{\ssM\ssN\ssP}$ there.

The integration over $A_{\ssM\ssN}$ is gaussian and so can be performed directly, leaving $V_{\ssM\ssN\ssP}$ as the dual field. Performing the gaussian integration requires an integration by parts, and so leaves a surface term
\be \label{stdef}
  L_{\rm st} = + \frac{1}{3!} \nabla_\ssQ \Bigl( \epsilon^{\ssM\ssN\ssP\ssQ\ssR\,\ssT} V_{\ssM\ssN\ssP} A_{\ssR\ssT} \Bigr) \,,
\ee
to which we return later. The saddle point relates the 4-form field strength, $F_{(4)} = \exd V_{(3)}$, to the 2-form $A_{(2)}$ as follows
\be
 \check A^{\ssM\ssN} = A^{\ssM\ssN} + \varepsilon \, e^{(r+1)\phi} \, Z^{\ssM \ssN} = - \frac{1}{2\cdot 3!} \,e^\phi \, \epsilon^{\ssM\ssN\ssP\ssQ\ssR\,\ssT} F_{\ssP\ssQ\ssR\,\ssT}  \,,
\ee
which inverts to
\be \label{eq:Fonshell}
 F_{\ssM\ssN\ssP\ssQ} = + \frac{1}{2} \, e^{-\phi} \, \epsilon_{\ssM\ssN\ssP\ssQ\ssR\,\ssT} \check A^{\ssR\ssT} \,.
\ee
The dual action (obtained after evaluating at this saddle point) then is
\bea \label{Lggenew}
 L_{\rm gge} &=& L_\ssZ + L_\ssA + L_{\rm mix} + L_{\rm lm} \nn\\
 &=& \frac14 \,\Lambda(\phi) \, Z_{\ssM \ssN}Z^{\ssM \ssN} - \frac{\varepsilon}{2\cdot 4!} \, e^{(r+1) \phi} \epsilon^{\ssM\ssN\ssP\ssQ\ssR\,\ssT} Z_{\ssM\ssN} F_{\ssP\ssQ\ssR\,\ssT}  \\
 && \qquad\qquad\qquad\qquad\qquad\qquad\qquad + \frac{1}{2 \cdot 4!} \, e^\phi \, F_{\ssP\ssQ\ssR\,\ssT} F^{\ssP\ssQ\ssR\,\ssT} +  L_{\rm st} \nn\\
 &=:& \check L_\ssZ + L_\BLF + L_\ssF + L_{\rm st} \,, \nn
\eea
where the last line defines $L_\BLF$ and $L_\ssF$, the latter of which also evaluates on shell to
\be
 L_\ssF := \frac{1}{2\cdot 4!} \, F_{\ssM\ssN\ssP\ssQ} F^{\ssM\ssN\ssP\ssQ} = - \check L_\ssA = - \frac12 \left( \frac{Q}{W^4} \right)^2 e^\phi\,.
\ee

Keeping in mind that $\cL_{\rm st} = \sqrt{-g} \; L_{\rm st}$ does {\em not} change the bulk equations of motion, in these variables the contribution of the Maxwell field to the RHS of the dilaton equation, \pref{dilatoneq}, becomes
\be
 \kappa^2 \, \frac{\partial L_{\rm gge}}{\partial \phi} = \kappa^2 \left[ \frac{\Lambda'}{\Lambda} \, \check L_\ssZ + (r+1) \, L_\BLF + L_\ssF \right]
\ee
instead of $\kappa^2 \left( - L_\ssA  + p L_\ssZ + r L_{\rm mix} \right)$. Similarly, since $\cL_\BLF = \sqrt{-g} \; L_\BLF$ is proportional to $Z_{(2)} \wedge F_{(4)}$ it does not couple to the metric at all, so the stress energy coming from $L_{\rm gge}$ is
\be
 T^{\ssM\ssN} = \frac{1}{3!} \,  F^{\ssM \ssA\ssB\ssC}{F^\ssN}_{\ssA\ssB\ssC} + \Lambda \, Z^{\ssM \ssA}{Z^\ssN}_{\ssA} - \Bigl(L_\ssF + \check L_\ssZ\Bigr) \, g^{\ssM \ssN}  \,,
\ee
and so contributes to $\cX$ as $\cX_{\rm gge} = -\frac12 \, {T^m}_m = -2 \check L_\ssZ + (L_\ssF + \check L_\ssZ) = L_\ssF - \check L_\ssZ$. Furthermore, the 6D trace becomes ${T^\ssM}_\ssM = 8 L_\ssF + 4 \check L_\ssZ - 6 \bigl(L_\ssF + \check L_\ssZ\bigr) = 2 \bigl(L_\ssF - \check L_\ssZ\bigr)$ and so the full 6D traced Einstein equation is
\bea \label{ReqF}
  \frac{\cR}{\kappa^2} &=& - 2 \Bigl( L_{\rm kin} + L_{\rm gm} \Bigr) -3  L_{\rm pot} + L_\ssF - \check L_\ssZ \nn\\
  &=& - 2 \Bigl( L_{\rm kin} + L_{\rm gm} + L_{\rm pot} + \check L_\ssZ - L_\ssF \Bigr) - \Bigl( L_{\rm pot} + L_\ssF - \check L_\ssZ \Bigr)\,.
\eea

If we define $\cY$ so that $\Box \phi = \kappa^2 \left( \cX + \cY \right)$ remains true, then we are led to replace \pref{Ydef} with
\be \label{Ydefnew}
  \cY = (q-1) V_b + \left(1 + \frac{\Lambda'}{\Lambda} \right) \check L_\ssZ + (r+1) \, L_\BLF  \,.
\ee
As expected, $L_\ssF$ drops out of this since the bulk Maxwell action does not break the scale symmetry.

Trading $A_{(2)}$ for $F_{(4)}$ in the surface term, $L_{\rm st}$, of \pref{stdef} allows it to be written
\be \label{stdef1}
  L_{\rm st} = + \frac{2}{4!} \, \nabla_\ssQ \Bigl( \epsilon^{\ssM\ssN\ssP\ssQ\ssR\,\ssT} V_{\ssM\ssN\ssP} A_{\ssR\ssT} \Bigr) = + \frac{1}{3!} \, \nabla_\ssQ \Bigl( V_{\ssM\ssN\ssP} \, e^\phi \, \check F^{\ssM\ssN\ssP\ssQ} \Bigr) \,,
\ee
where the second equality defines
\be
 \check F_{\ssM\ssN\ssP\ssQ} := F_{\ssM\ssN\ssP\ssQ} - \frac{\varepsilon}2  \, e^{r\phi}  \epsilon_{\ssM\ssN\ssP\ssQ\ssR\,\ssT} Z^{\ssR\ssT} \,.
\ee
Because the 4-form field equations imply $\nabla_\ssM \left( e^\phi \check F^{\ssM\ssN\ssP\ssQ} \right) = 0,$ evaluating $L_{\rm st}$ at a 4-form solution gives
\be
 \Bigl( L_{\rm st} \Bigr)_{\rm on-shell} = -\frac{1}{4!} \,  e^\phi \, \check F^{\ssM\ssN\ssP\ssQ} F_{\ssM\ssN\ssP\ssQ} = - 2L_\ssF - L_\BLF  \,,
\ee
and so on-shell the gauge action evaluates to
\be \label{6Dggeonshell}
 \Bigl( L_{\rm gge} \Bigr)_{\rm on-shell} = \check L_\ssZ + L_\BLF + L_\ssF +  L_{\rm st}
 = \check L_\ssZ - L_\ssF \,,
\ee
in agreement with the expected value, $L_{\rm gge} = \check L_\ssZ + \check L_\ssA$, in the original variables.

Although this dual formulation is equivalent to the original one, it makes several features usefully manifest. First, because $F_{(4)}$ turns out to be proportional to ${}^\star \check A_{(2)}$ rather than ${}^\star A_{(2)}$, it provides a natural way to express the change of variables from $A$ to $\check A$ directly in the action rather than the field equations, even for nontrivial dilaton profiles. In so doing it generates the same renormalization of the $Z$ kinetic term obtained earlier \cite{Companion}. Second, because the $V-Z$ coupling term has the form of $F_{(4)} \wedge Z_{(2)}$ it is immediate that this term is independent of the metric and so does not directly gravitate. This can also be understood in the original variables \cite{Companion}, in terms of a cancellation of localized contributions between $L_{\rm mix}$ and $L_\ssA$.

For the maximally symmetric configurations described above, evaluating $F_{(4)}$ using the solution to the $\check A_{(2)}$ field equation gives
\be \label{VvsQ}
 F_{\mu\nu\lambda\kappa} = e^{-\phi} \epsilon_{\mu\nu\lambda\kappa\rho\theta} \check A^{\rho\theta} = Q \, \check \epsilon_{\mu\nu\lambda\kappa}\,,
\ee
where $\check \epsilon_{\mu\nu\lambda\rho} = \pm \sqrt{- \check g}$ is the 4D volume form built from the metric $\check g_{\mu\nu}$. Notice that in the scale-invariant case (where $r = -1$) these definitions imply $F_{\mu\nu\lambda\kappa} \to s^2 F_{\mu\nu\lambda\kappa}$ under the scaling symmetry.

\subsection{Vortex solutions}
\label{sec:solutionswvortex}

This section describes an isolated vortex within a much larger ambient bulk geometry. Our goal is to establish that the presence of the dilaton couplings need not destroy the localized vortex solutions --- with exponentially falling solutions beyond the vortex radius, $\hat r_\varv$ --- familiar from the dilaton-free case. We also wish to relate the properties of the vortex to the asymptotic behaviour of the bulk fields and their derivatives outside of (but near to) the vortex itself, with a view to using these in the discussion of bulk solutions given earlier.

\subsubsection*{Nielsen-Olesen vortices}
\label{subsec:vortexsoln}

The isolated Abelian-Higgs system contains vortex solutions where the local gauge symmetry is relatively unbroken in a core region \cite{NOSolns,CStrings}, and the fields approach their vacuum solutions outside of this region. Although we consider a more complicated gravitating vortex sector that is coupled to a bulk scalar, as in $S_\ssV$, explicit numerical construction shows this only weakly perturbs the form of the localized vortex solutions in the parameter range of interest (as expected).

We work in a unitary gauge for which $\Psi = \psi$ is real (and for which the gauge fields $Z_\ssM$ and $A_\ssM$ are singular at the origin). We demand (as usual) the vortex fields approach their vacuum values away from the vortex, corresponding to $\psi \to v$ and $Z_\ssM \to 0$ far from the vortex core. Inside the vortex we have scalar boundary conditions $\psi(0) = 0$ and $\psi'(0) = 0$ in addition to gauge boundary condition $Z_\theta(0) = n_\varv /e$. This second boundary condition is chosen so that $Z$-flux quantization is satisfied within the vortex,
\be
 \oint\limits_{\rho = \rho_\varv} \d Z = 2 \pi \int\limits_{0}^{\rho_\varv} \d \rho \, \partial_\rho Z_\theta = 2 \pi \Bigl[Z_\theta(\rho_\varv) - Z_\theta(0) \Bigr] = - \frac{ 2 \pi n_\varv}{e} \,,
\ee
where $\rho_\varv$ is a point chosen sufficiently far from the vortex core that we can assume the gauge field takes on the vacuum value $Z_\theta = 0$, and the integer $n_\varv$ is the flux quantum of the vortex source in the region $X_\varv$ defined by $0 \le \rho \le \rho_\varv.$

\begin{figure}[t]
\centering
\includegraphics[width=\textwidth]{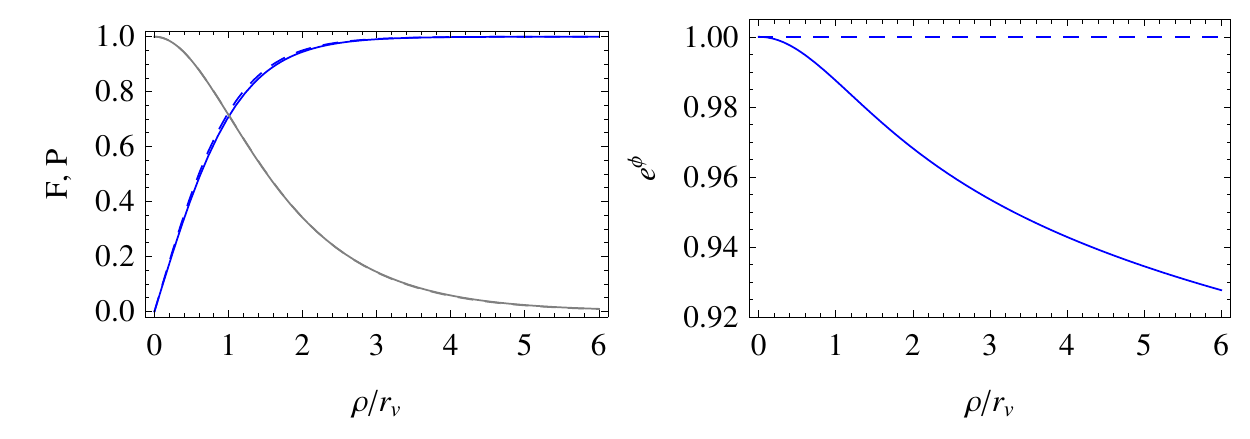}
\caption{These plots demonstrate that the dilaton couplings do not ruin the existence of localized vortex solutions. The left plot contains a comparison of the vortex profiles $F$ and $P$ for a non-gravitating vortex in flat space (dashed curves) and a gravitating vortex coupled to the dilaton (solid curves). The right plot shows the dilaton profile in the vortex region for both cases. The gravitating dilaton is slowly varying in the vortex region, with the change of $\phi$ over the vortex region being controlled by $\Delta e^{\phi} \sim \kappa^2 v^2$. The parameters used to generate the gravitating profiles are $d = 4$, $\varepsilon = 0.3$, $\beta = 3$, $\kappa v = 0.5$, $\phi(0) = 0$, $Q = 0.01 \, e v^2$, $V_0 = Q^2 / 2$ and the vortex sector is coupled to the dilaton through the generic choices $(p,q,r) = (-1,0,-1)$. The flat space profiles are generated by choosing $\kappa v = 0$ instead.}
\label{fig:profilecomparison}
\end{figure}

It is convenient when solving the vortex field equations to scale out the field dimensions by defining
\be
 Z_\theta = \frac{n_\varv P(\rho)}{e} \qquad \text{and} \qquad \psi = v F(\rho) \,.
\ee
In terms of these the boundary conditions become $F(0) = 0$ and $F \to 1$ far from the vortex, while $P(\rho)$ decreases from $P(0) = 1$ at the vortex core to its asymptotic value $P \to 0$. In these variables, the vortex field equations read as follows
\be
 \frac{1}{ B W^d} \left( B W^d F' \right)^\prime = \frac{n_\varv^2  P^2 F}{B^2 } + \lambda v^2 e^{q \phi} F (F^2 - 1 ) \,,
\ee
and
\be \label{Peqwderivterm}
\frac{1}{ B W^d} \left[ \Lambda(\phi) \left( \frac{W^d P' }{B}\right) \right]^\prime + (r+1) e^{(r+1) \phi} \left( \frac{e \varepsilon Q }{n_\varv} \right) \left( \frac{\phi'}{B W^d} \right) = \frac{e^2 v^2 F^2 P}{B^2} \,.
\ee
Examples of numerical solutions for the gravitating vortex profiles with dilaton interactions are shown in Fig.~\ref{fig:profilecomparison} and strongly resemble the nongravitating vortex solutions found in \cite{Companion} with constant dilaton in the vortex region $\phi = \phi_\varv$. In particular, they approach their asymptotic values exponentially over scales controlled by the $\phi_\varv$-dependent masses $m_\ssZ^2 =  \hat e^2(\phi_\varv) v^2 $ and $m_\Psi^2 = 2 \hat \lambda (\phi_\varv) v^2$ with
\be
 \hat e^2(\phi) := \frac{e^2 }{ \Lambda(\phi)} \qquad \hbox{and} \qquad
 \hat \lambda(\phi) := \lambda \, e^{q \phi} \,.
\ee

We wish to trace how the vortex solutions depend on $\phi$, and to do so it is useful to rescale factors of $\Lambda(\phi)$ into the coordinate $\rho$ (or, equivalently, the transverse metric), by writing
\be \label{eq:barcoords}
 g_{mn} \d y^m \d y^n = \Lambda(\phi) r_\varv^2  \left( \d \bar \rho^2 + \bar B^2 \d \theta^2 \right) = \hat r_\varv^2(\phi) \left( \d \bar \rho^2 + \bar B^2 \d \theta^2\right) \,.
\ee
With this choice the $\check A_{\bar\rho \theta}$ equation becomes
\be
 \partial_{\bar\rho} \left[ \frac{W^d e^{-\phi} \check A_{\bar\rho \theta}}{\bar B \Lambda(\phi)} \right] = 0 \,,
\ee
which has solution
\be
 \check A_{\bar\rho \theta} = \frac{\bar Q \bar B \, \Lambda(\phi) e^{\phi}}{W^d} \,,
\ee
with integration constant $\bar Q = Q \, r_\varv^2.$ The vortex field equations rewrite as
\be
 \frac{1}{\bar B W^d} \, \partial_{\bar \rho} \left[ \bar B W^d \partial_{\bar \rho} F \right] = \frac{n_\varv^2 P^2 F}{\bar B^2} + e^{q\phi }   \Lambda(\phi) \left( \frac{\lambda }{e^2 } \right) F(F^2-1) \,,
\ee
and
\be \label{Peqwderivtermbar}
 \frac{1}{\bar B W^d} \, \partial_{\bar \rho} \left( \frac{W^d \partial_{\bar \rho} P }{\bar B} \right) + (r+1) e^{(r+1) \phi} \left( \frac{e \varepsilon \bar Q }{n_\varv} \right) \left( \frac{\partial_{\bar\rho} \phi}{\bar B W^d} \right)  = \frac{F^2 P}{\bar B^2} \,.
\ee

Now comes the main point. To the extent that the gauge fields do not mix ({\em ie} $\varepsilon = 0$) or that the dilaton is approximately constant in the vortex region ($\partial_{\bar \rho} \phi \simeq 0$) these equations show that the vortex system depends on $\phi$ only through the one single, $\phi$-dependent parameter
\be
 \hat \beta(\phi) := e^{ q  \phi}\Lambda(\phi) \left( \frac{2 \lambda}{e^2 }\right)  = \frac{2\hat \lambda(\phi)}{\hat e^2(\phi)} = \frac{m_\Psi^2}{ m_\ssZ^2} = e^{ q  \phi}\Lambda(\phi) \beta \,.
\ee
Furthermore, if $p = 2r+1$ (as required if the $Z_\ssM$ kinetic energy is bounded below for all $\phi$) then the $\phi$-dependence of $\Lambda(\phi) = e^{p \phi }( 1 - \varepsilon^2)$ is simple and it is possible to make $\hat \beta$ independent of $\phi$ by choosing $p + q = 0$.

How important for these statements is the assumption that $\phi$ not vary across the vortex? Depedence on $\phi$ can enter if $\varepsilon$ is nonzero and $\phi$ actually does vary across the vortex, as can be seen from eq.~\pref{Peqwderivterm} or \pref{Peqwderivtermbar}. But because $\phi' \sim \kappa^2 v^2$ generically and $e \bar Q = e r_\varv^2 Q \sim (e/g_\ssA)(r_\varv/r_\ssB)^2$ the vortex system is only weakly sensitive to the value of $\phi$ for the one-parameter family $p = -q = 2r+1$, since the $\phi$-dependence arising from the $\phi'$ term of \pref{Peqwderivterm} is of relative size
\be
 \partial_{\bar \rho} \phi (r+1)  \varepsilon e \, r_\varv^2 Q \, e^{(r+1)\phi} \sim \kappa^2 v^2 (r+1) \left( \frac{\varepsilon e\, r_\varv^2}{g_\ssA r_\ssB^2} \right)\, e^{(r+1)\phi}  \sim (r+1) \left( \frac{\varepsilon g_\ssR^2}{e \, g_\ssA} \right) \, e^{(r+1)\phi} \,,
\ee
where the second estimate follows from the definition $r_\varv = 1/ev$ and $r_\ssB \sim \kappa/g_\ssR$. Because we are assuming $2r+1 = p$, the $\phi$-dependence can be absorbed by the hatted couplings to give
\be \label{KKsuppphi}
 \partial_{\bar \rho} \phi (r+1)  \varepsilon e \, r_\varv^2 Q \, e^{(r+1)\phi} \sim (r+1) \left( \frac{ \varepsilon \hat g_\ssR^2 }{\hat e \, \hat g_\ssA} \right) \,,
\ee
which is suppressed in the parameter range $\hat g_\ssA/ \hat e \approx \hat g_\ssR/\hat e \ll \kappa^2 v^2 \ll 1$ required to control the semiclassical approximation and to ensure $\ell \gg \hat r_\varv$.

\subsubsection*{BPS special case}

In the special case where $W = W_\varv$ and $\phi = \phi_\varv$ are constant in the vortex (for which a coordinate rescaling allows the choice $W_\varv = 1$), then the vortex field equations are the same as apply in the absence of the dilaton once we make the replacement $e \to \hat e$ and $\lambda \to \hat \lambda$ with $\hat e^2 := e^2/\Lambda(\phi_\varv)$ and $\hat \lambda := \lambda \, e^{q \phi_\varv}$. The vortex field equations in this case boil down to
\be \label{Peq}
 \frac{1}{B} \left( \frac{  P'}{ B} \right)' = \frac{\hat e^2 v^2 F^2 P}{B^2} \,,
\ee
while the $\psi$ equation becomes
\be \label{Feq}
 \frac{1}{B} \Bigl( B \; F' \Bigr)' = \frac{n_\varv^2 P^2 F}{B^2} + \hat \lambda v^2\, F\left( F^2 - 1 \right) \,.
\ee

The solutions to these equations are particularly simple when $\hat e^2 = 2 \hat \lambda$, since then eqs.~\pref{Peq} and \pref{Feq} are equivalent to the first-order equations,
\be \label{BPSeqs}
 B F' = n_\varv  F P  \qquad \hbox{and} \qquad
 \frac{n_\varv  P'}{\hat e B} = \sqrt{\frac{\hat \lambda}{2}} \; v^2  \left(  F^2 - 1 \right) \,.
\ee
We show later that $W = 1$ and $\phi = \phi_\varv$ also solve the bulk field equations when $\hat e^2 = 2 \hat \lambda$, and so this choice provides a consistent solution to all the field equations. Such solutions naturally arise when the vortex sector is itself also supersymmetric, since supersymmetry can require $\hat e^2 = 2 \hat \lambda$ and the vortices leave some supersymmetry unbroken.

When eqs.~\pref{BPSeqs} as well as $\phi = \phi_\varv$ and $W = 1$ hold, they also imply
\be
 L_{\rm kin} = \frac12 \, (\partial \psi)^2 =
  \frac{e^2}2 \, \psi^2 Z_\ssM Z^\ssM =  L_{\rm gm}\,,
\ee
and
\be \label{cLzeqVb}
 \check L_\ssZ := \frac14 \, \Lambda(\phi_b)  Z_{mn}Z^{mn} = \frac{\lambda}{4}  ( \psi^2 - v^2 )^2 = V_b \,,
\ee
which further imply that the vortex-localized contributions to $\cZ$ and $\cX$ cancel out: $\cZ_{\rm loc} = 0 $ and $ \cX_{\rm loc} = 0$. But the bulk contribution to $\cZ$ also vanishes if $\phi' = 0$ and --- as can be seen from eq.~\pref{newEinstein} --- it is the vanishing of $\cZ$ that allows constant $W$ to solve the Einstein equations. Finally, the dilaton field equation with constant $\phi$ requires $\check \cX_\ssB + \cY = 0$ everywhere, and so separately evaluating in the bulk and vortex implies $\check \cX_\ssB = \cY = 0$ separately. Although $\cY = 0$ can be ensured using the scale-invariant choices $p = r = -1$ and $q = 1$, vanishing $\check \cX_\ssB$ in general either requires a condition on the bulk gauge field, $Q$ (which need not agree with what is required by flux quantization) or a runaway to $e^\phi \to 0$.

Finally, the vortex part of the action evaluates in this case to the simple result
\be \label{eq:Tofv}
 S_\varv := \frac{1}{\sqrt{- \check g } } \int\limits_{X_\varv} \exd^2y \, \sqrt{-g} \; \varrho_{\, \rm loc}  = 2\pi \int\limits_0^{\rho_\varv} \exd \rho \, B \; \Bigl[ L_\Psi + V_b + \check L_\ssZ \Bigr] = \pi n_\varv v^2 \,,
\ee
where the second equality also defines the localized energy density $\varrho_{\, \rm loc} = L_{\rm loc}$.

\begin{figure}[h]
\centering
\includegraphics[width=\textwidth]{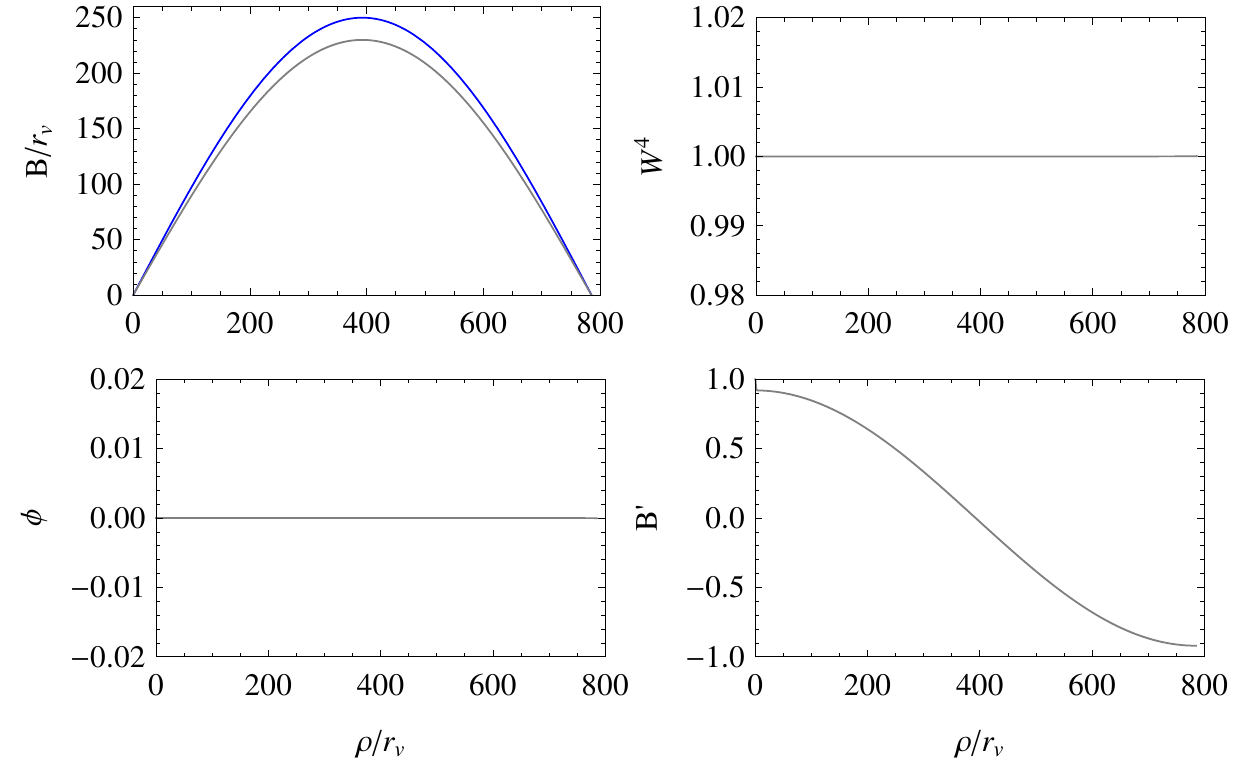}
\caption{These plots show the numerical bulk solution for $n=1$ BPS vortex sources ($\hat \beta = 1$) with scale invariant couplings to the dilaton $(p,q,r) = (-1,1,-1)$ and $\kappa v = 0.4.$~For these choices, the field equations are solved by a constant dilaton and warp factor: $W = 1$ and $\phi = 0.$ Because $\phi^\prime = W^\prime = 0$, this solution falls into the simple class of rugby ball solutions described in \S\protect\ref{ssec:bulksolns}.~In the top left plot the metric function $B$ is plotted against the same solution for a sphere of proper radius $\ell = 250 \pi r_\varv$, $B_{\rm sphere} = \ell \sin(\rho/\ell).$~The vortices (which cannot be resolved in these plots) introduce a defect angle into the bulk metric such that $B = \alpha \ell \sin(\rho/\ell)$ in the bulk. The defect angle, $\alpha$, can be determined by extrapolating $B'$ to the apparent singular point at $\rho_\star \approx 0.$ ~This yields $1 - \alpha \simeq 0.08 = \kappa^2 \check T / 2 \pi = (\kappa v)^2 /2 $ as expected from \protect\pref{eq:Tofv} and \protect\pref{dbGamma=derivsnew}. }
\label{fig:BPSsolution}
\end{figure}

\subsection{Integral relations}
\label{sec:integrals}

In this section we generalize the integral relations described earlier for the bulk system to include the vortex sources. Instead of integrating only over the exterior region, $\Bext$, we now instead integrate over the small regions, $X_\pm$, containing each vortex source, and thereby learn how the vortex determines the boundary conditions on the interface with $\Bext$. Using these boundary conditions for the integrated bulk solution is equivalent to integrating the bulk-vortex field equations over the entire space, $X_{\rm tot} := \Bext \cup X_+ \cup X_-$, which is smooth and compact and without boundary. In what follows we generically represent by $X_\varv = \{ X_+, X_-\}$ when we do not need to specify which source is of interest.

\subsubsection*{Integration over near-source pillboxes}

We first integrate just over $X_\varv $ to find the UV theory's perspective on the general matching conditions \cite{6DSUSYUVCaps, UVCaps} that relate near-source derivatives to properties of the source. These are useful in developing the effective theory of the next section that treats the vortices as point-like sources, or branes.

\pagebreak
\medskip\noindent{\em Maxwell fields}

\medskip\noindent
We start with the gauge-field equations. The simplest of these to solve is the Maxwell equation, \pref{Acheckeom}, since this does not depend directly on the fields $Z_\ssM$ or $\psi$. The solution is as before
\be \label{Acheckeomsoln}
  \check A_{\rho\theta} =  \frac{QB \, e^\phi}{W^d}  \,,
\ee
where $Q$ is an integration constant. This enters into the Einstein equations, \pref{avR4-v1}, \pref{avR2} and \pref{newEinstein}, through the combination $\check L_\ssA = \frac12 (Q/W^d)^2 \, e^\phi$.

Suppose now we take a test charge that couples only to $A_\ssM$ and ask how much flux it measures when taken around the vortex. We do so by moving around the edge of $V$, remaining everywhere outside the vortex. The edge of the pillbox is at a distance $\rho_\varv \gsim \hat r_\varv$ from the vortex so vortex fields are exponentially small, but we also choose $\rho_\varv \ll \ell$ so it contains a negligible fraction of the external bulk. The flux seen by this charge is

\bea \label{AfluxV}
 \Phi_\ssA(X_\varv) &:=& \int\limits_{X_\varv} A_{(2)} = \int\limits_{X_\varv} \Bigl( \check A_{(2)} - \varepsilon e^{(r+1)\phi} \, Z_{(2)} \Bigr) \nn\\
 &=& 2\pi \left[ Q \int\limits_0^{\rho_\varv} \exd \rho \; \frac{B  e^\phi}{W^d} - \varepsilon \int\limits_0^{\rho_\varv} \exd \rho \; e^{(r+1)\phi} Z_{\rho\theta} \right] \,.
\eea
The first term in the last equality gives the amount of bulk flux lying within $\rho < \rho_\varv$ in the absence of the vortex source, and so is negligibly small in the limit $\hat r_\varv \to 0$. The second term does survive this limit, however, because even though bulk fields vary slowly over the small vortex volume, the profile for $Z_{\rho\theta}$ is strongly peaked in such a way as to give the total quantized $Z$-flux,
\be
 \Phi_\ssZ = \int\limits_{X_\varv} Z_{(2)}
 \approx -\frac{2 \pi n_\varv}{e} \,.
\ee
So eq.~\pref{AfluxV} shows that the mixing of $A$ and $Z$ gauge fields through $L_{\rm mix}$ implies the test charge sees a vortex-localized component of flux despite it not coupling directly to the $Z$ field
\be
 \Phi_\ssA(X_\varv) \simeq - \varepsilon \, \Phi_\ssZ(X_\varv) \, e^{(r+1)\phi_\varv} \simeq   \frac{2\pi n_\varv \varepsilon }{e} \; e^{(r+1)\phi_\varv} \,,
\ee
where the approximation is true to the extent that $\hat r_\varv \ll \ell$. This localizes part of the external $A$ flux onto the vortex.

\medskip\noindent{\em Dilaton}

\medskip\noindent
Integrating the dilaton field equation, \pref{dilatoneom2}, over a vortex-containing pillbox gives
\be \label{phi'matching}
  \Bigl( BW^d \; \phi' \Bigr)_{\rho = \rho_\varv} = \frac{\kappa^2}{2 \pi} \int\limits_{X_\varv} \d^2 y  \sqrt{g_2} \, W^d \left( \cX + \cY \right) =  \frac{\kappa^2}{2\pi} \Bigl \langle \cX + \cY \Bigr\rangle_{\varv}   \,,
\ee
where we use that $BW^d \phi'$ vanishes at the vortex centre $\rho = 0$. We also repurpose the angle bracket notation of \pref{eq:anglebrack} to more general integration regions, $X_\varv$. This exact result expresses how the near-source limit of $\phi'$ just outside the vortex is determined by the detailed vortex profiles (keeping in mind that for both regions $X_{\pm}$, $\phi'$ in this expression is evaluated for a proper-distance coordinate for which $\rho$ {\em increases} as one moves away from the vortex).

If $\rho_\varv$ lies within the Kasner regime --- for which \pref{powerforms} applies --- the the left-hand side of \pref{phi'matching} becomes
\be \label{zGamma=derivs}
 \Bigl( BW^d \; \phi' \Bigr)_{\rho = \rho_\varv} = \left( \frac{z B_0 W_0^d}{\ell} \right) \left( \frac{\hat \rho_\varv}{\ell} \right)^{dw+b-1} \left[ 1 + \cO \left( \frac{\hat \rho_\varv}{\ell} \right) \right]
 = z \,\Gamma  \left[ 1 + \cO \left( \frac{\hat \rho_\varv}{\ell} \right) \right]  \,,
\ee
where $\hat \rho_\varv := \rho_\varv - \rho_\star$ and the second equality defines the quantity $\Gamma := B_0 W_0^d/\ell$ and uses the linear Kasner relation $dw+b=1$. When combined with \pref{phi'matching} this shows how vortex properties constrain combinations of bulk parameters (such as $z$) not already fixed by the bulk field equations.

\medskip\noindent{\em Metric}

\medskip\noindent
Similar conditions are obtained by integrating the Einstein equations over $X_\varv$. The trace-reversed Einstein equation, \pref{avR4-v1}, governing the curvature $\cR_{(d)}$ integrates to give
\be \label{dwGamma=derivs}
  dw \,\Gamma  \left[ 1 + \cO \left( \frac{\hat \rho_\varv}{\ell} \right) \right] = \left[  B \left( W^d \right)' \right]_{\rho_\varv}  = -\frac{1}{2\pi} \left[ \check R \, \left \langle W^{-2} \right\rangle_{\varv} + 2\kappa^2 \bigl\langle \cX \bigr\rangle_{\varv}\right] \,,
\ee
which uses the boundary condition $B \left( W^d \right)' = 0$ at $\rho = 0$ and rewrites $[B \left( W^d \right)']_{\rho_\varv}$ using \pref{powerforms}. This relates a different combination of bulk parameters to vortex properties. Integrating the $(\theta\theta)$ trace-reversed equation instead implies
\be \label{bGamma=derivs}
 b \, \Gamma  \left[ 1 + \cO \left( \frac{\hat \rho_\varv}{\ell} \right) \right]  = \left(  B' W^d \right)_{\rho_\varv} = 1  - \frac{\kappa^2}{2\pi} \left\langle \varrho - \cZ - \left( 1 - \frac{2}{d} \right) \cX \right\rangle_\varv  \,,
\ee
because of the boundary condition $B' W^d = 1$ at $\rho=0$.

Notice that the powers $z$, $w$ and $b$ are not independent since the bulk field equations imply they must satisfy the Kasner conditions \pref{Kasnerc}, and so the right-hand sides of the above expressions also cannot be completely independent in the limit $\hat \rho_\varv \ll \ell$. The resulting relations among the vortex integrals are developed in more detail in \S\ref{sec:6DIR} and play an important role in determining the off-brane components of the bulk stress energy in the effective theory applying at scales where the vortex size cannot be resolved. Because these relations follow from the bulk Einstein equations they can be regarded as general consequences of stress-energy conservation for the vortex integrals.

\subsubsection*{Integration over the entire transverse space}

We see that the integral relations of \S\ref{sec:bulksystem} give bulk properties in terms of near-vortex derivatives of bulk fields, and the integral relations just described then relate these near-vortex derivatives to explicit vortex integrals. The resulting relation between bulk properties and vortex integrals is more directly obtained by integrating over {\em all} of the transverse dimensions, $X_{\rm tot} = \Bext \cup X_+ \cup X_-$, at once. In such an integral all boundary terms cancel, as they must for any smooth compact transverse space. In the case of the Maxwell field integrating over the entire transverse space gives the flux-quantization condition, as discussed earlier.

\medskip\noindent{\em Dilaton}

\medskip\noindent
Integrating the dilaton field equation, \pref{dilatoneom2}, over the entire compact transverse dimension gives
\be \label{dilatontotinteq}
  \Bigl \langle \cX + \cY \Bigr\rangle_{\rm tot} = 0 \approx \bigl \langle \check\cX_\ssB \bigr\rangle_{\rm tot} + \sum_{\varv} \bigl \langle \cX_{\rm loc} + \cY \bigr\rangle_{\varv} \,,
\ee
where the approximate equality drops exponentially suppressed vortex terms when replacing a localized integral over the entire space with a localized integral over the source regions. Integration over the transverse space can be regarded as projecting the field equations onto the zero mode in these directions, and so \pref{dilatontotinteq} can be interpreted as the equation that determines the value of the dilaton zero-mode. (This conclusion is also shown more explicitly from the point of view of the effective $d$-dimensional theory in a forthcoming analysis \cite{STMicro}.) In the absence of the sources this zero mode is an exact flat direction of the classical equations associated with the scale invariance of the bulk field equations (for instance $\cX_\ssB = 0$ for the source-free Salam-Sezgin solution \cite{SS}) and the vortex contribution to \pref{dilatontotinteq} expresses how this flat direction becomes fixed when the sources are not scale-invariant.

\medskip\noindent{\em Metric}

\medskip\noindent
Integrating the trace-reversed Einstein equation over the entire transverse space leads to
\be \label{XthetathetaeEinsteininttot}
  \left\langle \varrho - \cZ - \left( 1 - \frac{2}{d} \right) \cX \right\rangle_{\rm tot} = 0 \,,
\ee
and
\be \label{intRdeq1}
  \check R \, \bigl\langle W^{-2} \bigr\rangle_{\rm tot} = - 2\kappa^2 \bigl\langle \cX \bigr\rangle_{\rm tot} = d\kappa^2  \bigl\langle L \bigr\rangle_{\rm tot} \,,
\ee
which shows how it is the stress-energy transverse to the source, $\langle \cX \rangle_{\rm tot} = -\frac{d}{2} \langle L \rangle_{\rm tot}$, that ultimately controls the size of the on-source curvature \cite{ScaleLzero}. Using \pref{dilatontotinteq} to eliminate $\cX $ from the right-hand-side similarly gives
\be \label{avR4-vdil}
 \check R \, \bigl\langle W^{-2} \bigr\rangle_{\rm tot} =  2\kappa^2 \bigl\langle \cY \bigr\rangle_{\rm tot} \approx \sum_{\varv} 2\kappa^2 \bigl\langle \cY \bigr\rangle_{\varv} \,,
\ee
whose approximation drops exponentially suppressed terms.

A final useful rewriting of these expressions uses the formula relating the $D$- and $d$-dimensional gravitational couplings, $\kappa^2$ and $\kappa_d^2$ respectively. Dimensionally reducing the Einstein-Hilbert action shows that this states
\be
 \frac{1}{\kappa_d^2} = \frac{1}{\kappa^2} \, \bigl\langle W^{-2} \bigr\rangle_{\rm tot} \,,
\ee
when the would-be zero-mode for $\phi$ is evaluated at the solution to its field equations, and so eqs.~\pref{XthetathetaeEinsteininttot}, \pref{intRdeq1} and \pref{avR4-vdil} at this point become
\be \label{R4-ddim}
 \check R = 2\kappa_d^2 \bigl\langle \cY \bigr\rangle_{\rm tot} = - 2\kappa_d^2 \bigl\langle \cX \bigr\rangle_{\rm tot} = - \left( \frac{2d}{d-2} \right) \kappa_d^2\,\bigl\langle \varrho  -  \cZ \bigr\rangle_{\rm tot} \,.
\ee

\section{Sources - effective IR description}
\label{sec:6DIR}

This section takes the point of view of a low-energy observer, and recasts the expressions for the UV theory found above into the language of the effective field theory appropriate in $D$ dimensions at scales much larger than the transverse vortex size, $\hat r_\varv$, but smaller than or of order the KK scale, $\ell$. We specialize for concreteness' sake to the case $(D,d) = (6,4)$, though our conclusions also hold more for general $D = d+2$.

\subsection{The EFT with point sources}

If the relevant length scale of an observable $r_{\rm obs}$ exceeds the length scale of the vortex, $r_{\rm obs} \gg  \hat r_\varv,$ then effects of the vortex can be organized as a series in the small quantity $\hat r_\varv / r_{\rm obs}$. For sufficiently large $r_{\rm obs}$, the internal structure of the vortex becomes irrelevant and it can be replaced with an idealized point-like object. This is why Abelian-Higgs vortices without localized flux are well-described by the Nambu-Goto string action at long distances \cite{NOSolns,CStrings}.

We here generalize this to include brane-localized flux, extending \cite{Companion} to include dilaton dependence. It is most convenient to do so using the dual formulation of the bulk action,
\be
 S_\ssB = - \int \d^6 x \sqrt{-g} \left[ \frac{1}{2 \kappa^2} \, g^{\ssM \ssN} \left( \cR_{\ssM \ssN} + \partial_\ssM \phi \partial_\ssN \phi \right) + \frac{ e^{\phi} }{2\cdot 4!} \left( F_{\ssM \ssN \ssP\ssQ} \right)^2 + \frac{2 g_\ssR^2}{\kappa^4} e^\phi  \right] \,,
\ee
and to include localized flux in the brane action we include the first subdominant term in a derivative expansion\footnote{The check here conforms with the notation of \cite{Companion} and is to remind that the use of the 4-form field automatically unmixes the gauge kinetic terms.}
\be \label{eq:Sbrane}
 \check S_\eff = -\sum_{\varv } \int_{x = z_\varv(\sigma)} \d^4 \sigma \sqrt{-\gamma} \left[ \check T_\varv(\phi) - \frac{1}{ 4!} \zeta_\varv (\phi) \, \varepsilon^{\mu\nu\lambda\rho} F_{\mu\nu\lambda\rho} \right] =: \sum_{\varv} \int\limits_{z_\varv} \exd^4 x \; \check \cL_\varv
 =: \int \exd^4 x \; \check  \cL_\eff \,,
\ee
where $\gamma_{\mu \nu}(\sigma) = g_{\ssM\ssN} \partial_\mu z_\varv^\ssM \partial_\nu z_\varv^\ssN$ is the induced metric at the position of the brane (with $z_\varv^\ssM(\sigma)$ denoting the brane position fields) and $\varepsilon^{\mu \nu\lambda \rho}$ is the totally antisymmetric 4-tensor associated with this metric. Since in what follows our interest is not in the dynamics of the brane position modes we assume a static vortex and choose coordinates so that it is located at fixed $y^m = y^m_\varv$ and identify $\sigma^\mu = x^\mu$ so $\gamma_{\mu\nu}(x) = g_{\mu\nu}(x,y_\varv) = W^{2}(y_\varv) \check g_{\mu \nu}(x)$. It is clear in both the UV and IR theories that the term linear in $F_{(4)}$ does not gravitate because it is metric independent, though this can also be inferred using the dual variables as in the dilaton-free case \cite{Companion}.

Because the effective theory cannot resolve the vortex structure it also cannot distinguish between the quantities $\rho_\varv$ and associated $\rho_\star$ used in previous sections. For each vortex we define the brane position in the effective theory to be the corresponding place where the external metric is singular when extrapolated using only bulk field equations. In practice this situates them at $\rho = \rho_\star$ in the coordinates used earlier, where $\rho_\star = \{\rho_\star^{+} , \rho_\star^- \}$ is determined by $B(\rho_\star) = 0$, so we use the notations $\phi_\star = \phi(\rho_\star) = \phi_\varv = \phi(y_\varv)$ interchangeably for the various bulk fields.

We show in this section that $\check S_\eff$ captures all of the physics of the full vortex action, up to linear order in the hierarchy $\hat r_\varv/\ell$, provided that the parameters $\check T_\varv(\phi)$ and $\zeta_\varv(\phi)$ are chosen appropriately. In general, these can be $\phi$-dependent quantities and we identify this dependence by demanding agreement between the predictions of $\check S_\eff$ and $\check S_\ssV$ to this order. Working to quadratic or higher order in $\hat r_\varv/\ell$ would require also including higher-derivative terms in $\check S_\eff$. To connect the point-brane action to the bulk fields we promote it to higher dimensions using a `localization' delta-function, $\delta(y)$. More precisely, we write
\be
 \check  S_\eff = \sum_\varv \int \d^\ssD x \; \check \cL_\varv \; \left( \frac{\delta(y-y_\varv)}{\sqrt{g_2} } \right) \,,
\ee
where $\delta(y)$ is localized around zero and it is normalized so that integrating over a single source region $X_\varv$ gives $\int_{X_\varv} \d^2 y \, \delta(y - y_\varv) = 1$. Performing the integration over the extra dimensions recovers the brane action in \pref{eq:Sbrane}.

As in \cite{Companion} a key question asks what fields the localizing function depends on since this affects how $\check S_\eff$ enters the bulk field equations. Although we assume in what follows that $\delta(y)$ is independent of the fields $A_{\ssM}$, $\phi$ and $g_{\mu \nu}$, we cannot also do so for the metric components $g_{mn}$ because it is designed to discriminate points based on proper distance from the vortex center. But because this metric dependence is only implicit it complicates the calculation of the brane's stress-energy components $\cT_{mn}$. One of the purposes of this section is to show how to determine these components without making ad-hoc assumptions about how $\delta(y)$ depends on $g_{mn}$, instead deducing them using properties of the bulk Einstein equations. Our conclusion ultimately is that the naive treatment of ignoring metric dependence in $\delta(y)$ need not be justified in the presence brane-dilaton couplings, and a cleaner way of inferring how a brane interfaces with the bulk is provided by directly relating the near-brane boundary conditions of bulk fields to derivatives of $\check \cL_\eff$ (along the lines of \cite{6DSUSYUVCaps, UVCaps}).

\subsection{Parameter matching}

We start by matching the coefficients\footnote{Although we often drop for simplicity the subscript `$\varv$' from $\check T$, $\zeta$ and $\rho_\star$ the reader should keep in mind that these quantities all can differ for different branes.} $\check T$ and $\zeta$ by comparing with the UV theory. A direct way to do so is by dimensionally reducing the UV action, and we verify that this also is what is required for $\check S_\eff$ to agree with $\check S_\ssV$ for observables like the components of the stress energy. When doing so it is crucial to notice that any such a comparison between the UV and IR theories need only be done up to linear order in $\hat r_\varv/\ell$ since it is only at this accuracy that the action \pref{eq:Sbrane} must capture the physics of earlier sections. This allows considerable simplification since integration of any slowly varying bulk quantity over $X_\varv$ vanishes quadratically with $\hat r_\varv$ for $\hat r_\varv \ll \ell$, allowing any such terms to be dropped to the accuracy with which we work. In what follows we accordingly take the formal limit $\hat r_\varv \to 0$ when discussing such integrals, by which we mean we drop terms that vanish at least quadratically in $\hat r_\varv/\ell$ in this limit. We take care {\em not} to similarly drop terms suppressed only by a single power of $\hat r_\varv/\ell$, however.

The term linear in $F_{\mu\nu\lambda\rho}$ in the UV action is given in \pref{Lggenew} as
\be
 L_\BLF = - \frac{\varepsilon}{2\cdot 4!} \, e^{(r+1) \phi} \epsilon^{\ssM\ssN\ssP\ssQ\ssR\,\ssT} Z_{\ssM\ssN} F_{\ssP\ssQ\ssR\,\ssT}
 = - \frac{\varepsilon}{2\cdot 4!} \, e^{(r+1) \phi} \epsilon^{mn\mu\nu\lambda\rho} Z_{mn} F_{\mu\nu\lambda\rho} \,,
\ee
and because the 4-form Bianchi identity, $\exd F_{(4)} = 0$, ensures the components $F_{\mu\nu\lambda\rho}$ cannot depend on the transverse coordinates, $y^m$, we know $F_{\mu\nu\lambda\rho}$ cannot be strongly peaked (unlike $Z_{mn}$) in the off-brane directions. Consequently integrating over the vortex area and comparing with the corresponding term in the IR theory gives
\be \label{zetamatch}
 \zeta_\varv(\phi_\varv) = \frac{\varepsilon}2 \lim_{\hat r_\varv \to 0} \int\limits_{X_\varv} \exd^2y \; B\, e^{(r+1) \phi} \, \epsilon^{mn} Z_{mn} \simeq -\left( \frac{ 2 \pi n_\varv \,\varepsilon }{e} \right) e^{ (r+1) \phi_\varv} \,,
\ee
where the approximate equality neglects the small variations of $\phi$ away from a constant value, $\phi_\varv$, at the vortex position, and uses $Z$-flux quantization to evaluate $\Phi_\ssZ(X_\varv) = \frac12 \int \exd^2 y \, \epsilon^{mn}Z_{mn} = - 2\pi n_\varv/e$.

To match the tension $\check T$ we compute the source contribution to the energy density. In the IR theory we have
\be
 \cT_{\mu \nu} = - g_{\mu \nu} \left[ V_\ssB + L_\phi - L_\ssF + \sum_\varv \left( \frac{\delta(y-y_\varv)}{\sqrt{g_2} } \right) \check T_\varv  \right]
\ee
and this is to be compared with the localized contribution to the stress energy of the UV theory integrated across the vortex,
\bea
 T_{\mu \nu} &=& \frac{1}{3!} \,  F^{\mu \ssM..\ssN}{F^\nu}_{\ssM..\ssN} - g_{\mu \nu} \Bigl( V_\ssB + L_\phi + L_\ssF + \check L_\ssZ  +  L_\Psi + V_b  \Bigr) \nn\\
 &=& - g_{\mu \nu} \Bigl( V_\ssB + L_\phi - L_\ssF + \check L_\ssZ  +  L_\Psi + V_b \Bigr) \nn\\
 &=& - g_{\mu \nu} \Bigl( \check \varrho_\ssB + \varrho_{\, \rm loc} \Bigr) \,.
\eea
Comparing $\cT_{\mu\nu}$ with $\lim_{\hat r_\varv \to 0} \langle T_{\mu\nu} \rangle_\varv$ reveals the localized contribution to the energy density in the UV theory to be (temporarily returning to general $d$)
\be \label{eq:Tmatching}
 W_\star^4 \, \check T [\phi(\rho_\star)] = \lim_{\hat r_\varv \to 0} \int\limits_{X_\varv} \exd^2y \, B W^4 \Bigl( \check L_\ssZ  + L_\Psi + V_b \Bigr) = \lim_{\hat r_\varv \to 0} \Bigl\langle \check L_\ssZ  + L_\Psi + V_b \Bigr\rangle_\varv = \lim_{\hat r_\varv \to 0} \langle \varrho_{\, \rm loc} \rangle_\varv \,.
\ee
We pause here to highlight one important feature of this result. The near-brane behaviour of the warp factor is $W \propto \rho^w$ and for $w > 0$ the quantity $W_\star^4 = W^4(\rho_\star)$ formally vanishes. The power law vanishing of $W$ can be reinterpreted as the logarithmic divergence of the field $c = \ln(W)$ and such divergences are common in theories with higher codimension brane sources. These divergences can be classically renormalized into the brane couplings \cite{6DHiggsStab,ClassRenorm} and in this case the tension would be renormalized such that the physical combination $W^4_\star \check T$ remains finite. The UV complete theory provides an explicit regularization of this divergence, since the vortex physics intervenes near the source to ensure $W > 0$ everywhere in the vortex region.

Unlike for $\zeta$, the result in \pref{eq:Tmatching} gives the $\phi$-dependence of $\check T$ only implicitly, so we next display this dependence more explicitly. We first compute how $\check T$ depends on $\phi$ assuming $\phi$ to be constant over a vortex. We expect the errors we make by doing so to be suppressed by powers of $\hat r_\varv/\ell$, and come back to verify this estimate shortly. The $\phi$-dependence of the tension is determined by the $\phi$-dependence of vortex integrals like
\bea
  \la V_b \ra_\varv = \frac14 \int\limits_{X_\varv} \exd^2y \; B W^4 \, \lambda(\phi)  \left( \psi^2 - v^2 \right)^2 \,, \quad&&\;
  \la \check L_\ssZ \ra_\varv = \frac14 \int\limits_{X_\varv} \exd^2y \; B W^4 \, \Lambda(\phi) Z_{mn} Z^{mn} \nn\\
  \la L_{\rm gm} \ra_\varv = \frac12 \int\limits_{X_\varv} \exd^2y \; B W^4 \, e^2(\phi) \psi^2 Z_m Z^m  \,, \quad&&\;
  \la L_{\psi\,{\rm kin}} \ra_\varv = \frac12 \int\limits_{X_\varv} \exd^2y \; B W^4 \, \partial_m \psi \partial^{\,m} \psi  \,,
\eea
and earlier sections show that the $\phi$-dependence of the vortex profiles appearing in these are fairly simple once expressed in terms of dimensionless variables, $F$ and $P$. Then the implicit $\phi$-dependence within the profiles themselves arises only through the combination $\hat\beta(\phi) = 2\hat \lambda(\phi)/\hat e^2(\phi)$, with $\hat \lambda(\phi) := \lambda \, e^{q\phi}$ and $e^2/\hat e^2(\phi) := \Lambda(\phi) = e^{p\phi} - \varepsilon^2 \, e^{(2r+1)\phi}$. We now ask whether any additional $\phi$-dependence arises from the integrations to set the scale of the above integrals.

To this end, it is useful to return to the variables used in \pref{eq:barcoords} in which $\bar B$ is dimensionless, as is the radial coordinate $\bar \rho.$ The integration measure $\d^2 y \, B W^4 = (1/ \hat e v)^2 \, \d^2 \bar y \, \bar B W^4$ and we have, for example,
\be
  \la V_b \ra_\varv \simeq \frac{\hat\lambda(\phi) v^2}{4\hat e^2(\phi)} \int\limits_{X_\varv} \exd^2 \bar y \; \bar B  W^4 \, \left( F^2 - 1 \right)^2
  = \hat\beta(\phi) \, \frac{v^2}{8} \int\limits_{X_\varv} \exd^2 \bar y \; \bar B  W^4 \, \left( F^2 - 1 \right)^2 \,.
\ee
Similarly
\be
  \la \check L_\ssZ \ra_\varv \simeq \frac{v^2}2 \int\limits_{X_\varv} \exd^2 \bar y \; \left(\frac{W^4 [P']^2}{\bar B} \right) \,, \quad
  \la L_{\rm gm} \ra_\varv \simeq \frac{v^2}2 \int\limits_{X_\varv} \exd^2 \bar y \; \left( \frac{W^4 F^2 P^2}{\bar B} \right) \,,
\ee
while
\be
  \la L_{\psi\,{\rm kin}} \ra_\varv \simeq \frac{v^2}2 \int\limits_{X_\varv} \exd^2 \bar y \; \bar B W^4 \, (F')^2 \,,
\ee
and so on. In these expressions the integrands are all proportional to $\hat e^2 v^4$, but the $\hat e^2$ dependence cancels the factors of $\hat r_\varv^2 = 1/(\hat e^2 v^2)$ coming from the integration measure to leave integrated results that again depend on $\phi$ only through $\hat \beta(\phi) \propto e^{q\phi} \Lambda(\phi)$.

\begin{figure}[t]
\centering
\includegraphics[width=\textwidth]{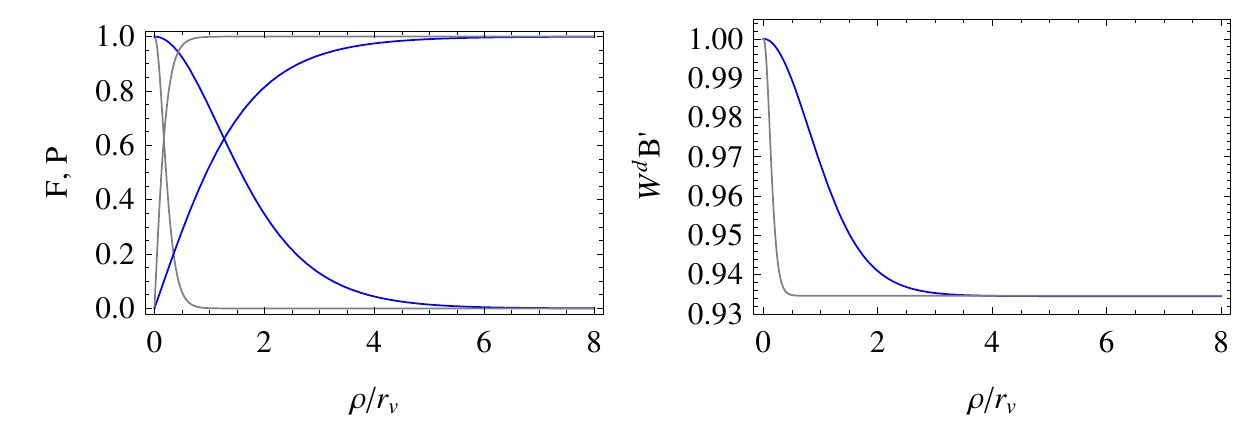}
\caption{A demonstration of the $\phi$-independence of $\check T_\varv$. Two solutions are presented for the decoupling choice $ p =  -q = 2r+1 = 2$ that differ only in the value of the dilaton in the vortex $\phi_\varv  \approx \phi(0)  = \{0,-2\}.$ The light (grey) lines represent the solution for the choice $\phi(0) = -2$ and the blue lines represent the solution for $\phi(0) = 0$. While the physical mass scales that control the size of the profiles, $m_\Psi^2(\phi_\varv) = 2 \check \lambda v^2  e^{q \phi_\varv} $ and $m_\ssZ^2(\phi_\varv) = e^{- p \phi_\varv} e^2 v^2 / ( 1 - \varepsilon^2)$, are demonstrably heavier for $\phi_\varv \simeq -2$ (since these profiles fall off much faster), the defect angle $\left[ W^d B ' \right]_{\rho_\varv} \simeq 1 - \frac{\kappa^2 \check T_\varv}{2 \pi}$ that measures the tension of the vortex is independent of $\phi_\varv$. The other parameters of this solution are $d=4$, $\varepsilon=0.6$, $\beta = 0.8$, $\kappa v = 0.4$, $Q=1.6 \times 10^{-4}$ and $V_0 = Q^2/2.$}
\label{fig:Tofphi}
\end{figure}

Consequently
\be
 \check T(\phi) = \check T[\hat \beta(\phi)] \,,
\ee
and in particular $\check T$ is $\phi$-independent for the one-parameter family of choices $p = -q = 2r+1$. Because $\zeta(\phi) \propto e^{(r+1)\phi}$ having {\em both} $\check T$ and $\zeta$ be $\phi$-independent happens only in the special case of scale invariance, for which $p = r = -q = 1$. This inference of the $\phi$-independence of the tension is verified numerically, as seen in Fig.~\ref{fig:Tofphi}.

What happens once we drop the assumption that $\phi'$ is negligible within the vortex? In this case our earlier estimate --- eq.~\pref{KKsuppphi} --- of the leading  $\phi$-dependence of vortex profiles implies that the leading $\phi$-dependent corrections to quantities like $\check T_\varv$ and $\langle \cX_{\rm loc} \rangle_\varv$ have the KK-suppressed form
\be \label{tX estimates}
 \frac{\delta \check T_\varv }{\check T_\varv} \approx \frac{\delta \langle \cX_{\rm loc} \rangle_\varv }{\langle \cX_{\rm loc} \rangle_\varv} \approx (r+1) \left( \frac{g_\ssR}{e } \right) e^{(r+1) \phi_\varv} \approx (r+1) \left(\frac{\hat g_\ssR}{\hat e} \right)  \approx (r+1) \kappa v \left(\frac{\hat r_\varv}{\ell} \right) \,,
\ee
which uses $g_\ssA \sim g_\ssR$ and $p=2r+1$ as well as $\hat r_\varv \sim 1/(\hat e v)$ and $\ell \sim \kappa/\hat g_\ssR$.

\subsection{Near-source matching conditions}

As noted earlier the hidden dependence of the localization function $\delta(y)$ on $g_{mn}$ complicates the inference of how branes contribute to the bulk field equations. In this section we give a $\delta$-independent way of expressing the bulk-brane interaction wherein knowledge of the brane action, $\check S_\eff$, directly gives the near-brane asymptotic derivatives of the bulk fields. This connection is the analogue in the IR theory of the expressions in \S\ref{sec:integrals} relating the near-vortex derivatives of bulk fields to integrals over the vortex. The agreement between the UV and IR descriptions of these boundary conditions provides a check on the matching between the two versions of the theory. The discussion found here also parallels the dilaton-free case \cite{Companion} and this source can be consulted for more details.

\medskip\noindent{\em Dilaton}

\medskip\noindent Since the general discussion is simplest for a scalar field we treat the dilaton first. In the effective theory of point-like branes the field equation for the dilaton reads
\be
 \frac{1}{\kappa^2 } \Box \phi = V_\ssB - L_\ssA + \sum_\varv \left( \frac{\delta(y-y_\varv)}{\sqrt{g_2} } \right) \left( \check T_\varv^\prime(\phi) - \frac{1}{4!} \zeta^\prime_\varv(\phi) \epsilon^{\mu \nu \lambda \rho} F_{\mu \nu \lambda \rho} \right)  \,.
\ee
As before we integrate this equation over a source-containing pillbox to isolate the near-brane derivative of the dilaton, recognizing that this pillbox can be taken to have infinitesimal size in the effective theory (within which the vortex size cannot be resolved). On the left-hand side the result is
\be
 \lim_{\hat r_\varv \to 0} \la \Box \phi \ra_\varv = 2 \pi \lim_{\rho \to \rho_\star} \Bigl( B W^d \phi^\prime \Bigr) = 2 \pi z \, \Gamma \,,
\ee
where the last equality uses the near-brane asymptotic bulk solution \pref{powerforms} and again uses the definition $\Gamma := B_0 W_0^d/\ell$. Performing the same operation on the right hand side gives
\be
 \lim_{\hat r_\varv \to 0} \left \langle V_\ssB - L_\ssA + \frac{ \partial \check  L_\eff}{\partial \phi} \right \rangle_\varv = W^d(\rho_\star) \left( T_\varv^\prime(\phi) - \frac{1}{4!} \zeta^\prime_\varv(\phi) \epsilon^{\mu \nu \lambda \rho} F_{\mu \nu \lambda \rho} \right)_{\rho = \rho_\star} \,,
\ee
where the smoothness of the bulk sources ensures their integral does not survive the limit $\hat r_\varv \to 0$. The localized sources do survive this limit, however, and give the final result
\be \label{zmatchform}
  z \, \Gamma = \lim_{\rho \to \rho_\star} \Bigl( B W^d \phi^\prime \Bigr) = \frac{\kappa^2 W^d_\star}{2\pi}  \left( T_\varv^\prime(\phi) - \frac{1}{4!} \zeta^\prime_\varv(\phi) \epsilon^{\mu \nu \lambda \rho} F_{\mu \nu \lambda \rho} \right)_{\rho_\star} = - \frac{\kappa^2}{2\pi \sqrt{-\check g}} \left( \frac{\delta \check S_\eff}{\delta \phi} \right)_{\rho=\rho_\star} \,,
\ee
in agreement with \cite{6DSUSYUVCaps, UVCaps}. Eq.~\pref{zmatchform} is useful because it directly extracts the impact of the brane-dilaton coupling on the bulk dilaton without reference to the localization function $\delta(y)$.

This argument shows it is useful to divide quantities like $\varrho$, $\cX$ and $\cZ$ into a localized piece, whose integral survives the point-like limit, and a smooth bulk piece that does not. For instance: $\cX = \check \cX_\ssB + \cX_{\rm loc}$, where the bulk part, $\check \cX_\ssB := V_\ssB - \check L_\ssA$, depends only on bulk fields and so satisfies $\langle \check \cX_\ssB \rangle_\varv \to 0$ as $\hat r_\varv \to 0$. The same need not be true for vortex-localized quantities like $\cX_{\rm loc}  = V_b - \check L_\ssZ$, $\cY$ and so on.

The analogous relation between the near-vortex derivative of the dilaton and the vortex sources in the UV theory is given by \pref{phi'matching}, which we rewrite for convenience of comparison here
\be \label{zmatchform2}
  \lim_{\rho \to \rho_\star}  B W^d \phi^\prime =  \frac{\kappa^2}{2\pi}  \lim_{\hat r_\varv \to 0} \Bigl \langle \cX_{\rm loc} + \cY \Bigr\rangle_{\varv} = \frac{\kappa^2}{2\pi} \lim_{\hat r_\varv \to 0}  \left\langle q V_b + \frac{\Lambda^\prime}{\Lambda} \check L_\ssZ + (r+1) L_{\BLF}\right\rangle_\varv \,.
\ee
Comparing with this UV version allows a check the consistency of our inference of the $\phi$-dependence of $\check L_\eff$. Comparing terms linear in $F_{(4)}$ gives
\be
 \zeta_\varv^\prime(\phi) = (r+1) \lim_{\hat r_\varv \to 0}  \frac{\varepsilon}{2} \int\limits_{X_\varv} \d^2 y \, B\,  e^{(r+1) \phi} \, \epsilon^{mn} Z_{mn} \,,
\ee
which is consistent with the earlier result \pref{zetamatch}, assuming the dilaton is approximately constant in the region $X_\varv$ and provided differentiation with respect to $\phi$ is performed with fixed vortex fields. Similarly comparing the 4-form-independent terms gives
\be
 \left( W^d \, \check T_\varv^\prime \right)_{\rho = \rho_\star} = \lim_{\hat r_\varv \to 0} \left \langle q V_b + \frac{\Lambda^\prime}{\Lambda} \check L_\ssZ \right \rangle_\varv \,,
\ee
which is also consistent with the earlier expression \pref{eq:Tmatching}.

\medskip\noindent{\em Metric}

\medskip\noindent A similar argument relates near-source metric derivatives to properties of the brane action. As above one integrates the Einstein equations over a region $X_\varv$ enclosing the vortex and finds two kinds of terms that survive the limit $\hat r_\varv \to 0$ of vanishingly small vortex size. One such class of terms comes from the vortex parts of the stress energy while the other come from terms inside the Einstein tensor involving second derivatives with respect to $\rho$, with all other contributions not singular enough to survive the small-vortex limit.

The simplest equation to analyze is the $(\rho\rho)$ Einstein equation, \pref{hamconstraint}, since this involves no second derivatives at all. Consequently its integral over $X_\varv$ simply states
\be \label{Trhorhovan}
 {\cT^\rho}_\rho = \lim_{\hat r_\varv \to 0} \langle {T^\rho}_\rho \rangle_\varv
 = \lim_{\hat r_\varv \to 0} \langle \cZ_{\rm loc} - \cX_{\rm loc} \rangle_\varv \simeq 0 \,.
\ee
Physically, this is a consequence of dynamical equilibrium for the non-gravitational microphysics of which the vortex is built, since this requires there to be no net radial pressure.

For all of the remaining Einstein equations the Einstein tensor does include second derivatives and so their integration over $\cX_\varv$ leads to the following relation between the near-source derivatives of the metric and the metric derivative of the source action
\be \label{eq:metricmatching}
 2\pi \lim_{\hat r_\varv \to 0} \Bigl[ \sqrt{-g}  \left( K^{ij}  - K g^{ij} \right) \Bigr]^{\rho_\varv}_0 = - \kappa^2 \sqrt{-g} \; \cT^{ij} = - 2 \kappa^2  \frac{\delta \check S_\eff }{\delta g_{ij}} \,,
\ee
where the derivative on the right-hand side is with respect to the metric evaluated at the brane position and $K_{ij}$ is the extrinsic curvature for surfaces of constant $\rho$ (with $K = g^{ij}K_{ij}$). The indices $i$ and $j$ run over all coordinates but $\rho$.

The derivative on the right-hand-side can be taken reliably for on-brane components of the metric $(ij) = (\mu \nu)$ using the action \pref{eq:Sbrane} and gives
\be
 \frac{\delta \check S_\eff }{\delta g_{\mu \nu} } = -\frac{1}{2} \sqrt{-\gamma } \, \check T \, g^{\mu \nu} \,.
\ee
For the metric ansatz of interest the extrinsic curvature components are $K_{\mu \nu} = W W' \check g_{\mu \nu}$ and $K_{\theta \theta} = B B'$, so the on-brane components of \pref{eq:metricmatching} give
\be \label{dbGamma=derivsnew}
  \lim_{\hat r_\varv \to 0} \left[  1 -  B W^4 \left(  (d-1) \frac{ W' }{ W} + \frac{ B' }{ B } \right)  \right]_{\rho = \rho_\varv} = \lim_{\rho \to \rho_\star} \left[ 1 - \left( \frac{d-1}{d} \right) B \left(W^d \right)^\prime - B^\prime W^d \right] = \frac{ \kappa^2 W_\star^4 \check T }{2 \pi} \,.
\ee
This result is to be compared with the appropriate linear combination of \pref{bGamma=derivs} and \pref{dwGamma=derivs} in the UV theory, keeping only those vortex-localized terms that survive in the limit $\hat r_\varv \to 0$:
\be \label{uvbprime}
 \lim_{\rho \to \rho_\star} \left[ 1 - \left( \frac{d-1}{d} \right) B \left( W^d \right)^\prime - B' W^d \right] \simeq \frac{\kappa^2}{2 \pi} \lim_{\hat r_\varv \to 0} \la \varrho_{\, \rm loc} + \cX_{\rm loc} - \cZ_{\rm loc} \ra_\varv  \simeq \frac{\kappa^2}{2 \pi} \lim_{\hat r_\varv \to 0} \la \varrho_{\, \rm loc} \ra_\varv \,.
\ee
The last equality here uses \pref{Trhorhovan}, leaving a result consistent with our earlier identification in eq.~\pref{eq:Tmatching} that $W_\star^d \check T_\varv \simeq \lim_{\hat r_\varv \to 0} \langle \varrho_{\, \rm loc} \rangle_\varv$.

\subsection{Brane and vortex constraints}

We now turn to the remaining case, where $(ij) = (\theta \theta)$ in \pref{eq:metricmatching}, which gives
\be \label{troublewith}
  \lim_{\hat r_\varv \to 0} \Bigl[ B \left(W^d \right)^\prime \Bigr]_{\rho_\varv} =
  \lim_{\rho \to \rho_\star} B \left(W^d\right)^\prime = \frac{\kappa^2}{\pi \sqrt{-\check g}} \, g_{\theta \theta} \left( \frac{\delta \check S_\eff }{\delta g_{\theta \theta} } \right) \,.
\ee
The trouble with this expression is that the dependence of $\check S_\eff$ on $g_{\theta\theta}$ is only known implicitly, so we cannot perform the differentiation on the right-hand side to learn about the near-brane derivatives on the left-hand side. Fortunately we may instead read this equation in the other direction: it tells us the right-hand side because the radial constraint -- eq.~\pref{hamconstraint} -- already determines the derivatives on the left-hand side in terms of known quantities. It is ultimately this observation that allows us to determine $\langle \cX_{\rm loc} \rangle_\varv$ and $\langle \cZ_{\rm loc} \rangle_\varv$ separately in terms of the quantities $\check T$ and $\zeta$ \cite{6DSUSYUVCaps,UVCaps}.

To see this we first compare with the corresponding UV expression, using \pref{dwGamma=derivs} to rewrite
\be \label{dwGamma=derivsnew}
  \lim_{\rho \to \rho_\star}   B \left( W^d \right)'  \simeq  - \frac{\kappa^2}{\pi} \lim_{\hat r_\varv \to 0}  \bigl\langle \cX_{\rm loc} \bigr\rangle_\varv \,,
\ee
and so
\be
 {\cT^\theta}_\theta = \frac{2}{\sqrt{-\check g}} \, g_{\theta \theta} \left( \frac{\delta \check S_\eff }{\delta g_{\theta \theta} } \right) = - 2 \lim_{\hat r_\varv \to 0}  \bigl\langle \cX_{\rm loc}  \bigr\rangle_\varv \simeq - \lim_{\hat r_\varv \to 0}  \bigl\langle \cX_{\rm loc} + \cZ_{\rm loc} \bigr\rangle_\varv  \,,
\ee
which confirms that the vortex stress-energy component ${\cT^\theta}_\theta$ captures the integral of the vortex-localized part of ${T^\theta}_\theta$ in the UV theory.

To fix ${\cT^\theta}_\theta$ we follow \cite{6DSUSYUVCaps,UVCaps} and again use the constraint equation \pref{hamconstraint}, but rather than integrating it over the vortex we instead evaluate it at $\rho = \rho_\varv$, just outside the vortex, and solve it for $B \left( W^d \right)'$ to find
\be \label{reconstraint}
 \left( \frac{d-1}{d} \right) B \left( W^d \right)' = - \left( B^\prime W^d \right) + \left( B^\prime W^d \right) \sqrt{ 1 + \left( \frac{d-1}{d} \right) E } \,,
\ee
where we define
\be \label{eq:EV}
 E := \left( \frac{B}{B^\prime} \right)^2 \left[ 2 \kappa^2 \left( \cZ_\ssB - \check \cX_\ssB\right)  - W^{-2} \check R \right] \,.
\ee
Because this gives $B \left( W^d \right)'$ in terms of $B' W^d$ and other known quantities we use it in \pref{troublewith} to get an explicit expression for ${\cT^\theta}_\theta$ (and so also for $\langle \cZ_{\rm loc} \rangle_\varv \simeq \langle \cX_{\rm loc} \rangle_\varv$).

So far eq.~\pref{reconstraint} assumes only that $\rho_\varv$ is sufficiently far from the vortex that localized contributions to $\cX$ and $\cZ$ are exponentially small. However, since $\rho_\varv \sim \hat r_\varv$ in size we can also drop terms quadratically small in $\rho_\varv$ when comparing with the point-brane theory. Since both the $\check R$ and $\check \cX_\ssB$ terms in \pref{eq:EV} are proportional to $B^2(\rho_\varv) \sim \rho^2_\varv \sim \hat r_\varv^2$ they can be dropped in this limit. By contrast, the term containing $ {2 \kappa^2} \cZ_\ssB(\rho_\varv) =  [\phi^\prime(\rho_\varv) ]^2$ need not vanish quadratically in this limit, since the asymptotic power-law form \pref{powerforms} allowed in this region implies
\be
 \lim_{\hat r_\varv \to 0} E \simeq \lim_{\rho \to \rho_\star} 2 \kappa^2 \left( \frac{B}{B^\prime} \right)^2 \cZ_\ssB \simeq \lim_{\rho \to \rho_\star} \left( \frac{B \phi^\prime }{B^\prime} \right)^2 = \left( \frac{z}{b} \right)^2 \,,
\ee
and the final equality uses \pref{zGamma=derivs} and \pref{bGamma=derivs} to rewrite the right-hand side in terms of Kasner powers. Indeed, using this final result in \pref{reconstraint} and trading the remaining terms for Kasner powers leads to
\be
 ( d-1 ) w  \simeq - b + b \sqrt{1 + \left( \frac{d-1}{d} \right) \left( \frac{z}{b} \right)^2 }  \,,
\ee
which reveals it not to be independent in this limit of the two Kasner conditions, \pref{Kasnerc}.

\begin{figure}[t]
\centering
\includegraphics[width=\textwidth]{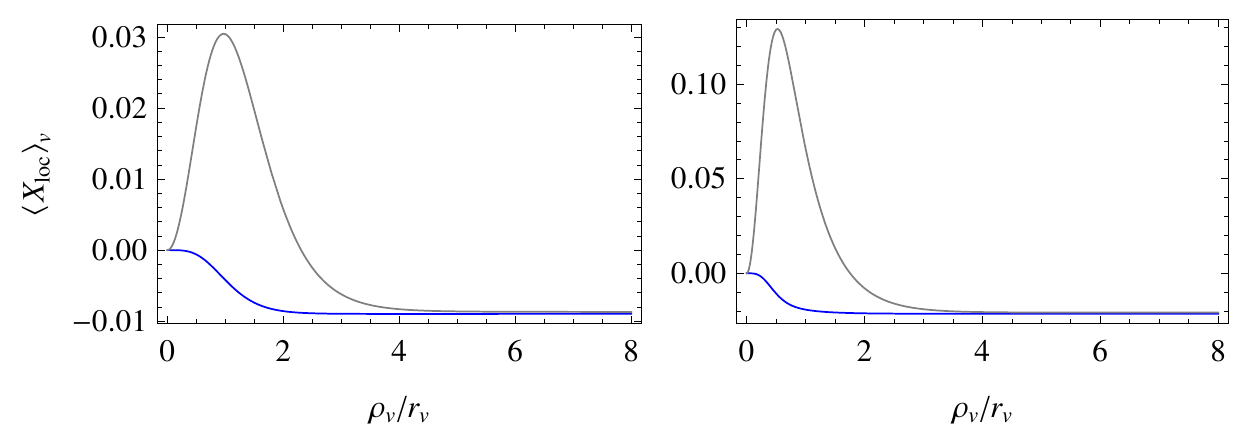}
\caption{Two examples of the vortex constraint \protect\pref{eq:vortexconstraint}. The light (grey) line is calculated from the expression $\kappa^2 \la \cX_{\rm loc} \ra_\varv \simeq - \pi \left[ B (W^d)^\prime \right]_{\rho_\varv}$ and the dark (blue) line is calculated using using the second brane constraint. Once evaluated outside of the vortex $\rho_\varv \gsim \hat r_\varv \approx r_\varv $, the two independently calculated quantities are in perfect agreement, and their failure to agree inside the region $\rho_\varv \lsim \hat r_\varv$ reflects the fact that the vortex constraint does not hold locally in the source. In the first plot the parameters are $d=4$, $\varepsilon=0$, $\beta = 3$, $\kappa v = 0.5$, $\phi_\varv \approx \phi(0) =0$, $Q=2\times 10^{-4}$ and $V_0 = Q^2/2.$ The vortex is coupled to the dilaton via the choice $(p,q) = (0,2)$ and $r$ is not relevant because gauge kinetic mixing has been shut off. In the second plot, the only change in parameters is that $\phi(0) = 1$. }
\label{fig:vortexconstraint}
\end{figure}

In passing we pause to remark on a point already alluded to in earlier sections: that the constraints imposed by the Einstein equations also imply the existence of relations among vortex integrals --- like $\langle \cX_{\rm loc} \rangle_\varv$ and $\langle \cY \rangle_\varv$ --- for {\em arbitrary} vortex microphysics in the point-source limit. The constraint is found by eliminating $B'$ and $W'$ in terms of vortex integrals using the near-vortex expressions \pref{bGamma=derivs} and \pref{dwGamma=derivs}, leading for instance to
\be
 \lim_{\hat r_\varv \to 0} E \simeq \left( \frac{z}{b} \right)^2 \simeq  \left( \frac{\kappa^2 \la \cX_{\rm loc} + \cY \ra_\varv }{1 - \frac{\kappa^2}{2 \pi} \la  \varrho_{\, \rm loc} - 2 \cX_{\rm loc} \left( \frac{d-1}{d} \right) \ra_\varv } \right)^2  \,.
\ee
In limit of small $\kappa^2 \langle \cY \rangle_\varv$ and $\kappa^2 \langle \cX_{\rm loc} \rangle_\varv$ the resulting constraint simplifies to
\be \label{eq:vortexconstraint}
 \kappa^2 \la \cX_{\rm loc} \ra_\varv \approx - \frac{\kappa^4 \la \cY \ra_\varv^2}{ 4 \pi} \,.
\ee
This brane constraint can be verified using explicit numerical solutions, as is illustrated in Fig.~\ref{fig:vortexconstraint}. The suppression it implies for $\kappa^2 \langle \cX_{\rm loc} \rangle_\varv$ relative to $\kappa^2 \langle \cY \rangle_\varv$ is the analog of the suppression found earlier, \pref{eq:kasnersolved}, between the various Kasner exponents that imply $w$ and $1-b$ are order $z^2$ when $z \ll 1$. This is seen most explicitly by dividing \pref{zGamma=derivs} by \pref{dwGamma=derivs} and evaluating in the $\hat r_\varv \to 0$ limit, which gives
\be \label{weakconstraint}
 \frac{2 \la \cX_{\rm loc} \ra_\varv} {\la \cX_{\rm loc} + \cY \ra_\varv} \simeq - \frac{d w}{z} \approx - \frac{z}{2} + \cO(z^2)
\ee
where the approximation follows from \pref{eq:kasnersolved}. The integrated vortex sources $\langle \cX_{\rm loc} \rangle_\varv$ and $\langle \cY \rangle_\varv$ cannot adjust independently to leading order in $\delta$ and $\hat r_\varv / \ell$. This is explored in more detail in the next subsection.

To summarize, it is possible to be very explicit about the dependence on $\check S_\eff$ of $\kappa^2 \langle \cX_{\rm loc} \rangle_\varv$ and $\kappa^2 \langle \cZ_{\rm loc} \rangle_\varv$ in the limit when these vortex integrals are small. In this case $\langle \cY \rangle_\varv$ is given by differentiating $\check S_\eff$ by eqs.~\pref{zmatchform} and \pref{zmatchform2},
\be \label{zmatchform3}
  z \, \Gamma = \lim_{\rho \to \rho_\star} \Bigl( B W^d \phi^\prime \Bigr) = - \frac{\kappa^2}{2\pi \sqrt{-\check g}} \left( \frac{\delta \check S_\eff}{\delta \phi} \right)_{\rho=\rho_\star} \simeq  \frac{\kappa^2}{2\pi}  \lim_{\hat r_\varv \to 0} \la \cY \ra_{\varv} \,,
\ee
Then $\langle \cZ_{\rm loc} \rangle_\varv$ and $\langle \cX_{\rm loc} \rangle_\varv$ are obtained from \pref{Trhorhovan} and \pref{eq:vortexconstraint}.

This section also shows how errors can arise when treating the localizing function $\delta(y)$ as a naive delta-function that is independent of of $g_{mn}.$ Such a treatment would mistakenly conclude that the brane action in \pref{eq:Sbrane} is independent of the bulk metric, and give
\be
 \lim_{\rho \to \rho_\star} B (W^d)^\prime = \frac{\kappa^2}{\pi \sqrt{-\tilde g} } g_{\theta \theta} \left( \frac{\delta \check S_{\rm eff}}{\delta g_{\theta \theta} } \right) = 0  \qquad \text{(naive result)} \,.
\ee
This not consistent with \pref{dwGamma=derivsnew} unless the vortex source has vanishing $\langle \cX_{\rm loc} \rangle_\varv$ in the limit $\hat r_\varv \to 0,$ which is not necessarily true, since the vortex must satisfy the simplified vortex constraint \pref{eq:vortexconstraint}. In the next section we estimate this quantity's (often nonvanishing) size.

\section{The scale of the response}
\label{sec:gravresponse}

This section summarizes the implications of the previous sections for the generic size of back-reaction effects on physical quantities. In particular we broadly scope out how the size of the on-vortex curvature, $\check R$, varies with the parameters $p$, $q$ and $r$ governing the size of the vortex-dilaton couplings.

As \pref{R4-ddim} suggests, the $d$-dimensional curvature is sensitive to $p$, $q$ and $r$ through the size of $\kappa_d^2 \langle \cY \rangle_{\rm tot}$. We now trace how the size of $\check R$ can drastically change as these parameters take various limits. The size of the quantity $\langle \cX_{\rm loc} \rangle_\varv$ also varies strongly in these limits since it is also related to $\langle \check \cY \rangle_{\rm tot} \approx \sum_\varv \langle \check \cY \rangle_\varv$ because of \pref{Trhorhovan}.

\subsection{Generic case}

Before examining special cases we first establish a baseline by considering the generic case. Recall from the definition of $\cY$ in the dual variables,
\be
 \cY = (q - 1 ) V_b + \left( 1 + \frac{\Lambda^\prime}{\Lambda} \right) \check L_\ssZ + (r+1) L_{\BLF} \,,
\ee
that each term contains localized vortex fields and for generic choices of parameters these all integrate over a vortex to give
\be
 \langle \cY \rangle_\varv \sim \langle \varrho_{\, \rm loc} \rangle_\varv \sim \check T \sim v^2 \,,
\ee
much as we saw earlier.

When this is true, it follows from the vortex constraints \pref{eq:vortexconstraint} and \pref{Trhorhovan} that the vortex integrals of the transverse stress-energy components are suppressed relative to $\langle \cY \rangle$ by an additional factor of $\kappa^2 v^2$,
\be
 \kappa^2 \la \cZ_{\rm loc} \ra_\varv \simeq \kappa^2 \la \cX_{\rm loc} \ra_\varv \sim \kappa^4 v^4 \,.
\ee
This suppression is perhaps not surprising given that these quantities vanish in the flat space limit, as discussed in Appendix~\ref{app:scaling}, as a consequence of stress-energy balance.

In this generic case eq.~\pref{zmatchform2} then implies the following size for the near-source dilaton derivative
\be
 \lim_{\hat r_\varv \to 0} \left [B W^d \phi^\prime \right]_{\rho_\varv} \simeq \frac{\kappa^2}{2 \pi} \la \cX_{\rm loc} + \cY \ra_\varv \sim \frac{\kappa^2}{2 \pi} \la \cY \ra_\varv \sim \kappa^2 v^2 \,,
\ee
with a similar size predicted by \pref{uvbprime} for the near-brane form of $B'$ (and so also the brane defect angle)
\be \label{B'resultx}
 1 - \lim_{\hat r_\varv \to 0} \left ( B' W^d \right)_{\rho_\varv} \simeq \frac{\kappa^2}{2 \pi} \la \varrho_{\, \rm loc} \ra_\varv \sim \kappa^2 v^2 \,.
\ee
By contrast, \pref{dwGamma=derivsnew} predicts the near-brane limit of $W'$ is additionally suppressed,
\be
 \lim_{\hat r_\varv \to 0}  \left [B (W^d)^\prime \right]_{\rho_\varv} \simeq - \frac{\kappa^2}{\pi}\la \cX_{\rm loc} \ra_\varv \sim \kappa^4 v^4  \,.
\ee
All of these estimates are confirmed by the numerical results displayed in Fig.~\ref{fig:bigkasner} and Fig.~\ref{fig:vortexconstraint}, and they are consistent with the weakly gravitating solutions to the Kasner equations (which state that $1-b$ and $dw$ are order $z^2$ when $z \ll 1$).

The $d$-dimensional curvature in this generic case is not particularly small, since it is of order
\be
 \check R = 2 \kappa^2_d \la \cY \ra_{\rm tot} \sim \kappa^2_d v^2 \,,
\ee
corresponding to a vacuum energy of order $v^2$.

\subsection{Scale invariance}

The other extreme is the scale invariant case: $(p,q,r) = (-1,1,-1)$. As noted previously, in this case the quantity $\cY$ everywhere vanishes,
\be
 \cY = 0 \,,
\ee
and so the vortex constraints, \pref{eq:vortexconstraint} and \pref{Trhorhovan}, then also imply
\be
 \la \cZ_{\rm loc}\ra_\varv \simeq \la \cX_{\rm loc} \ra_\varv \simeq 0\,.
\ee
This in turn leads to the a vanishing near-brane limits for both $\phi'$ and $W'$,
\be
 \lim_{\hat r_\varv \to 0} \left [B W^d \phi^\prime \right]_{\rho_\varv} \simeq \frac{\kappa^2}{2 \pi} \la \cX_{\rm loc} \ra_\varv \simeq 0
 \qquad \hbox{and} \qquad
 \lim_{\hat r_\varv \to 0} \left [B (W^d)^\prime \right]_{\rho_\varv} \simeq - \frac{\kappa^2}{\pi}\la \cX_{\rm loc} \ra_\varv \simeq 0 \,,
\ee
although the brane defect angle is again given by \pref{B'resultx} and so is not particularly suppressed.

This suppression of $W'$ in the near-brane limit resembles the pure Maxwell-Einstein (dilaton-free) case considered in \cite{Companion}, for which the radial Einstein constraint also generically forces $\langle \cX_{\rm loc} \rangle_\varv$ and the near-brane limit of $W'$ to vanish. It is also consistent with the observation that, in the scale invariant case, there exists the BPS choice of couplings $\hat \beta = 1$ for which $\cX_{\rm loc} = \cZ_{\rm loc} = 0$ locally and to all orders in $\hat r_\varv / \ell$. This is particularly obvious in the numerical BPS solution of Fig.~\ref{fig:BPSsolution} which has constant warp factor and dilaton.

Lastly, the scale-invariant choice ensures the vanishing of the $d$-dimensional curvature
\be
 \check R = 2 \kappa^2_d \la \cY \ra_{\rm tot} = 0 \,,
\ee
although this is less interesting than it sounds since this is typically achieved by having the dilaton zero mode run away, $e^\phi \to 0$, since this is the only generic solution to the zero-mode equation $0 = \langle \cX + \cY \rangle_{\rm tot} = \langle \cX \rangle_{\rm tot} \propto e^\phi$.

\subsubsection*{Special case: supersymmetric `BPS' branes}

The coefficient of the scale invariant runaway potential vanishes if the flux quantization condition can be satisfied such that $Q=Q_{\rm susy}$, where
\be \label{QequalsQSUSY}
Q_{\rm susy} := \pm \frac{2g_{\ssR}}{\kappa^2} \,.
\ee
In other words, $\langle \cX \rangle_{\rm tot}$ can be made to vanish in a $\phi$-independent way if this relation holds, because it ensures that $\check \cX_\ssB = V_\ssB- L_\ssA = 0$ by having the individual contributions cancel against one another locally. The localized contributions to $\langle \cX \rangle_{\rm tot}$ also vanish locally if $\check \beta = 1$ and are otherwise suppressed.

In the scale invariant case, for which $\phi^\prime = 0$ at the source, we can insert this desired value of $Q = Q_{\rm susy}$ into the the flux quantization condition in \pref{Fluxqnappl}, and derive a local condition relating tension and localized flux. For $N = +1$ it reads
\be \label{eq:bpsbrane}
\Phi_{\ssA\pm} = \frac{\pi}{g_\ssR}(1-\alpha_\pm) \quad\implies\quad \kappa^2 \check T_\pm = - 2g_\ssR \zeta_\pm
\ee
so long as $\alpha_+ = \alpha_-$. Otherwise, the defect angles contribute in a nonlinear way to the flux quantization condition via $\Upsilon := \sqrt{\alpha_+\alpha_-}$ and no such local condition can be found. Note also that the relation \pref{QequalsQSUSY} can only be made to hold if the gauge symmetry has the coupling strength of the $R$-symmetry, $g_\ssA = g_\ssR$.

Such a relation between the defect angle and flux is also expected from the supersymmetry conditions, in order to guarantee the continued cancellation of spin and gauge connections within the Killing spinor's covariant derivative, once localized sources are introduced \cite{AccidentalSUSY,DistributedSUSY}. However, past works did not properly identify the defect angle with the renormalized brane tension, as in \pref{eq:bpsbrane}.

\subsection{Decoupling case: $p  = -q = 2r+1$}

The `decoupling' choice $p = -q = 2r+1$ is special because it ensures that $\hat \beta$ is $\phi$-independent. This suppresses the dilaton-dependence of the brane tension, as in \pref{KKsuppphi}, which also makes it similar in form to the $\phi$-dependence of the brane-localized flux, $\zeta$. Notice both become $\phi$-independent in the scale-invariant special case where $r = -1$.

We here ask whether this suppression of brane-dilaton couplings also suppresses the brane's gravitational response, and if so by how much. To decide, recall that for the decoupling choice, $\cY$ takes on the special form
\be
 \cY = (q-1) ( V_b - \check L_\ssZ ) + (r+1) L_{\BLF} = (q-1) \cX_{\rm loc} + (r+1) L_{\BLF} \,,
\ee
and the vortex constraint in \pref{eq:vortexconstraint} becomes
\be
 \kappa^2 \la \cZ_{\rm loc} \ra_\varv \simeq \kappa^2 \la \cX_{\rm loc} \ra_\varv \simeq -\frac{\kappa^2}{ 4 \pi} \,  \la q \cX_{\rm loc} + (r+1) L_{\BLF} \ra_\varv^2
 \simeq - \frac{1}{4\pi} \, (r+1)^2 \kappa^4 \la  L_{\BLF}\ra^2_\varv   \,,
\ee
which uses that \pref{eq:vortexconstraint} requires $\langle \cX_{\rm loc} \rangle_\varv \ll \la  L_{\BLF}\ra_\varv$.

Conveniently, the quantity $\la L_{\BLF}\ra_\varv$ can be estimated using the on-shell value of the 4-form field strength \pref{eq:Fonshell} and the bulk gauge field strength \pref{Acheckeomsoln}. This gives
\be
 \la L_{\BLF}\ra_\varv =  \varepsilon Q \int \d \rho \, Z_{\rho \theta} \,  e^{(r+1) \phi} \simeq - Q \left( \frac{ 2 \pi n_\varv \varepsilon}{e} \right) e^{(r+1)\phi_\varv}  \,,
\ee
where $n_\varv$ and $\phi_\varv$ are the flux quantum and approximately constant value of the dilaton in in the vortex region $X_\varv$. To get a handle on its size we use the source free estimate $Q \sim \pm g_\ssR/\kappa^2$ and write the result more transparently in terms of $\hat r_\varv^{-1} = \hat e(\phi) v = e v \, e^{-p\phi/2}$ and $\ell^{-1} = 2 \hat g_\ssR(\phi)/\kappa = (2g_\ssR/\kappa) e^{-\phi/2}$, to find
\be
 \kappa^2 \la L_{\BLF}\ra_\varv \simeq - 2 \pi n_\varv \varepsilon \left( \frac{\kappa^2 Q }{e} \right) e^{(r+1)\phi_\varv}
 \approx \mp 4 \pi n_\varv \varepsilon \left( \frac{\hat g_\ssR(\phi_\varv) }{\hat e(\phi_\varv)} \right)
 = \mp 4 \pi n_\varv \varepsilon \, \kappa v  \left( \frac{\hat r_\varv }{\ell} \right) \,.
\ee
For small vortices, $\hat r_\varv/\ell \ll 1$, this reveals the decoupling case to lie between the generic and scale-invariant cases, with $\kappa^2 \langle \cY \rangle_\varv$ suppressed by a single power of the vortex size in KK units, as might be expected given that the leading $\phi$-dependence of the point brane action arises within the single-derivative localized-flux term of the point-brane action.

These estimates lead to the following expectations for near-brane bulk derivatives. As always, $B'$ near the sources is dominated by the energy density, with
\be
 1 - \lim_{\hat r_\varv \to 0} \left[  B' W^d \right]_{\rho_\varv} \simeq \frac{\kappa^2}{2\pi} \la \varrho_{\, \rm loc} \ra_\varv \,,
\ee
while (by contrast) the near-source derivatives $\phi'$ and $W'$ are KK suppressed
\be
 \lim_{\hat r_\varv \to 0} \left [B W^d \phi^\prime \right]_{\rho_\varv} \simeq \frac{\kappa^2}{2 \pi} \la \cY \ra_\varv \simeq \mp (r+1) 2 n_\varv \varepsilon \, \kappa v \left( \frac{\hat r_\varv }{\ell} \right)  \,,
\ee
and
\be
 \lim_{\hat r_\varv \to 0} \left [B (W^d)^\prime \right]_{\rho_\varv} \simeq - (r+1)^2 \left( 2 n_\varv \varepsilon \, \kappa v \right)^2 \left( \frac{\hat r_\varv }{\ell} \right)^2 \,.
\ee
Because it is suppressed by two powers of $\hat r_\varv/\ell$ this last expression is of the same size as terms we have neglected, such as second-derivative terms in the brane action, so we should not trust its precise numerical prefactor.

The $d$-dimensional curvature in this case can be written similarly to give $\check R = 2 \kappa^2_d \la \cY \ra_{\rm tot}$, whose size corresponds to an effective vacuum energy, $U_\eff$, of order
\be \label{Ueffexpression}
 U_\eff \approx \la \cY \ra_{\rm tot} \simeq \mp 4 \pi n_\varv \varepsilon  \kappa v  \left( \frac{\hat r_\varv }{\ell} \right) \; \frac{1}{\kappa^2} \,.
\ee
Strictly speaking, this expression is somewhat self-referential because it is a function of the would-be dilaton zero-mode, $\varphi$, whose value must be obtained by minimizing a quantity like \pref{Ueffexpression}. Although this generically leads to runaway behaviour in the scale-invariant case (with $U_\eff \propto e^{2\varphi}$ implying the minimum occurs for $\varphi \to -\infty$, the same need not be true when $r \ne -1$ since the vortex action then breaks scale invariance and so changes the functional form of $U_\eff(\phi)$.

What the above expressions leave open is what this precise form is, since this requires a more detailed evaluation of the $\phi$-dependence of all quantities that has been done here, including all of the $\phi$-dependence implicit within $\langle \cX \rangle_{\rm tot}$, and not just the near-source part $\langle \cX_{\rm loc} \rangle_\varv$ estimated here. Although this takes us beyond the scope of this (already long) paper, we do describe such a more detailed calculation in \cite{STMicro}, including a description of the 4D perspective obtained by integrating out the extra dimensions entirely.

\section{Discussion}
\label{section:discussion}


This paper's aim has been to carefully determine the way in which codimension-two objects back-react on their in environment, for a specific UV completion which captures the physics of brane-localized flux coupled to the bulk dilaton. To this end, we have determined the way in which the microscopic details of the vortex get encoded in IR observables, such as the size of the transverse dimensions and the on-brane curvature.

Quite generally, we find that the breaking of scale invariance in the vortex sector leads to modulus stabilization in the IR and --- for the particular `decoupling' choice of couplings $p=-q=2r+1$ --- we find a parametric suppression in the value of the on-brane curvature, by a single power of the small ratio $\hat r_\varv/\ell$. What remains to be determined is whether reasonable choices for vortex-dilaton couplings can stabilize the extra dimensions with a sufficiently large hierarchy, $\ell/\hat r_\varv$, to profit from this suppression. We explore this in more detail in a companion paper \cite{STMicro}, where we find such a stabilization to be possible.

This UV completion verifies earlier claims \cite{BLFFluxQ, 6DSUSYUVCaps, 6DHiggsStab} that moduli can be stabilized in a codimension-two version of the Goldberger-Wise mechanism \cite{GoldWis}, when the dilaton couplings of the vortex are chosen to break scale invaria

This work leaves many open questions. One such asks what the effective description is for the dilaton dynamics in the theory below the KK scale within which the extra dimensions are integrated out. In particular how does such a theory learn about flux quantization, which we've seen is central to the dynamics that stabilizes the extra dimensions. We also address this question in the companion work \cite{STMicro}.

Another open direction asks whether vortex configurations can be contrived that break supersymmetry in a {\em distributed} way (as proposed in \cite{NatSmallCC}, for example, with some supersymmetry unbroken everywhere locally but with all supersymmetries broken once the entire transverse space is taken into account). One might hope to construct a locally half-BPS UV vortex --- using, {\em eg}, a configuration of hyperscalars as in \cite{Swirl} --- and embed two (or more) of them in the bulk in such a way that leaves supersymmetry completely broken globally.

We leave these questions for future work.

\bigskip
\begin{center}
NOTE ADDED
\end{center}

\medskip\noindent The preprint \cite{deltas} appeared shortly after our posting of this preprint. As we describe in more detail in our reply \cite{reply}, we believe its use of $\delta$-function techniques makes it insufficiently precise to resolve questions about the size of the vacuum response for 6D systems. It is insufficiently precise due to its implicit assumption that the $\delta$-function is independent of the bulk fields; an assumption that is most suspicious for the transverse metric, given that the $\delta$-function is designed to discriminate points based on their proper distance from the brane sources. A proper statement of the bulk field dependence requires a more precise regularization (and renormalization) of the brane action, as given for instance in \cite{Companion, UVCaps} and this paper. The upshot of these more precise procedures is that it is stress-energy balance (as expressed through the radial Einstein `constraint' equation) that determines any hidden dependence on bulk fields, and this is easy to get wrong if one uses a naive regularization for which the regularization radius is not set dynamically using a microscopic set of field equations.

\section*{Acknowledgements}

We acknowledge Florian Niedermann and Robert Schneider for collaborations at early stages of this work as well as discussions about this paper and \cite{Companion} while they were in preparation. We thank Ana Achucarro, Asimina Arvanitaki, Savas Dimopoulos, Gregory Gabadadze, Ruth Gregory, Mark Hindmarsh, Stefan Hoffmann, Leo van Nierop, Massimo Porrati, Fernando Quevedo, and Itay Yavin for useful discussions about self-tuning and UV issues associated with brane-localized flux. The Abdus Salam International Centre for Theoretical Physics (ICTP), the Aspen Center for Physics, the Kavli Institute for Theoretical Physics (KITP), the Ludwig-Maximilian Universit\"at, Max-Planck Institute Garsching and the NYU Center for Cosmology and Particle Physics (CCPP) kindly supported and hosted various combinations of us while part of this work was done. This research was supported in part by funds from the Natural Sciences and Engineering Research Council (NSERC) of Canada, and by a postdoctoral fellowship from the National Science Foundation of Belgium (FWO), by the Belgian Federal Science Policy Office through the Inter-University Attraction Pole P7/37, the European Science Foundation through the Holograv Network, and the COST Action MP1210 `The String Theory Universe'. Research at the Perimeter Institute is supported in part by the Government of Canada through Industry Canada, and by the Province of Ontario through the Ministry of Research and Information (MRI). Work at KITP was supported in part by the National Science Foundation under Grant No. NSF PHY11-25915. Work at Aspen was supported in part by National Science Foundation Grant No. PHYS-1066293 and the hospitality of the Aspen Center for Physics.

\appendix

\section{Scaling and the suppression of $\langle \cX_{\rm loc}\rangle$}
\label{app:scaling}

We here derive a useful integral identity that is satisfied by the vortex solutions in the limit where gravitational back-reaction is neglected so the vortex is in flat space. It is this identity that underlies the small size of vortex integrals like $\langle \cX_{\rm loc} \rangle$ encountered in the main text.

The starting point is the observation that the static vortex solution minimizes the energy (or negative action)
\bea
 I &=& \int \exd^2 y \sqrt{-g} \Bigl(L_\phi + L_{\Psi} + V_b +  L_\ssA + L_\ssZ + L_{\rm mix} \Bigr) \nn\\
 &=& 2\pi \int_0^\infty  \sqrt{-g}\left[ \frac{1}{2\kappa^2} \, g^{mn} \partial_m \phi \, \partial_n \phi + V_\ssB(\phi)  \right. \nn\\
 && \qquad\qquad  +\frac12 \, g^{mn} \Bigl( \partial_m \Psi \partial_n \Psi + e^2 \Psi^2 Z_m Z_n \Bigr)  + \frac{\lambda}4 \,  e^{q\phi} \left(\Psi^2 - v^2 \right)^2 \nn\\
 && \qquad\qquad\qquad\qquad \left. + \frac14 \, e^{-\phi} A_{mn} A^{mn}  + \frac14 \, e^{p\,\phi} Z_{mn} Z^{mn}  + \frac12 \, \varepsilon  e^{r\phi} Z_{mn} A^{mn} \right] \,,
\eea
and observes that this is stationary with respect to arbitrary variations of the matter fields (without also varying the metric), due to their field equations. In particular it is invariant under rescalings of the form $\phi(y) \to \phi(\sqrt{s}\; y)$, $\Psi(y) \to \Psi(\sqrt{s}\; y)$ and so on.

Now, suppose the metric $g_{mn}$ also satisfies $g_{mn}(\sqrt{s}\; y) = s g_{mn} (y)$, such as is true for the case of a locally flat metric, $g_{mn} \, \exd y^m \exd y^n = \exd \rho^2 + \alpha^2 \rho^2 \, \exd \theta^2$ under the rescaling $\rho \to \sqrt{s}\; \rho$. Since $I$ is invariant under {\em arbitrary} coordinate transformations, $y^m \to \xi^m(y)$, in the 2D directions, provided {\em both} the matter fields and metric transform, the stationarity of $I$ with respect to redefinitions $\phi(y) \to \phi(\sqrt{s}\;  y)$ (for all matter fields) is equivalent (for 2D metrics with conformal Killing vectors) to stationarity with respect to the rescaling $g_{mn} \to s g_{mn}$ without also performing the coordinate rescaling.

Under the rescaling $g_{mn} \to s \, g_{mn}$ we have $\sqrt{-g} \to s \sqrt{-g}$ and so (assuming all fields vary and point only in the transverse dimensions)
\bea
 I &=& \int \exd^2 y \sqrt{-g} \Bigl( L_\phi + L_{\Psi} + V_b + L_\ssA + L_\ssZ + L_{\rm mix} \Bigr) \nn\\
 &\to& \int \exd^2 y \sqrt{-g} \left[ L_{\phi} + L_{\Psi} + s \Bigl( V_\ssB + V_b \Bigr) + \frac{1}{s} \Bigl( L_\ssA + L_\ssZ + L_{\rm mix} \Bigr) \right]  \,,
\eea
and so the stationarity condition becomes
\bea \label{virial}
 0 = \left( \frac{\exd I}{\exd s} \right)_{s=1} &=&  \int \exd^2 y \sqrt{-g} \left[  \Bigl( V_\ssB + V_b \Bigr) - \Bigl( L_\ssA + L_\ssZ + L_{\rm mix} \Bigr) \right] \nn\\
 &=&  \int \exd^2 y \sqrt{-g} \left[  \Bigl( V_\ssB + V_b \Bigr) - \Bigl( \check L_\ssA + \check L_\ssZ \Bigr) \right]  \nn\\
 &=&  \int \exd^2 y \sqrt{-g} \; \cX \,.
\eea
The claim is that this equation is an automatic consequence of the matter equations of motion, and expresses the balancing of pressures (on average) in the radial directions for a stable vortex configuration.

The same arguments apply equally well for an isolated $Q=0$ vortex for which $L_\phi$ and $\check L_\ssA$ are negligible in $I$, in which case eq.~\pref{virial} reduces to
\be \label{virialspcase}
 0 =   \int \exd^2 y \sqrt{-g} \Bigl( V_b  - \check L_\ssZ \Bigr)  =   \int \exd^2 y \sqrt{-g} \; \cX_{\rm loc} \,.
\ee
In the special BPS case examined in the body this is not only true on average but is also locally true, following directly from eq.~\pref{cLzeqVb}.

Notice that the way it has been derived shows that eq.~\pref{virial} is a statement about the vanishing of the extra-dimensional components of the stress energy, as is made more explicit in the appendix of \cite{Companion}. It need not hold once the metric back-reaction is turned on, but the vanishing of the flat-space result leads to the exact result being is smaller than might otherwise have been expected.

\end{document}